%% file: main.tex
\documentclass[11pt]{article} 
\usepackage{iclr2025_conference,times}
\usepackage{titlesec}

\usepackage{CJKutf8}
\usepackage{booktabs}
\usepackage{enumitem}

\usepackage{amssymb}
\usepackage{amsmath}
\setlist[itemize]{itemsep=5pt,topsep=5pt}

\usepackage{mdframed}

\usepackage{caption}
\usepackage{url}
\usepackage{enumitem}
\usepackage{lipsum}

\usepackage[authordate,legalnotes=false,dashed=false,backend=biber,noibid, hyperall=true, maxcitenames=2, uniquelist=false, uniquename=false]{biblatex-chicago}

\DeclareDelimFormat{nameyeardelim}{\addcomma\space}

\renewbibmacro*{cms:shorthandintro}{%
  \iffieldundef{shorthand}
    {}%
    {\printtext[brackets]{\printfield{shorthand}}}%
}

\makeatletter
\renewcommand{\@makefnmark}{\hbox{\@textsuperscript{\tiny\@thefnmark}}}
\makeatother

\bibliography{appendix_refs,references}

\usepackage{tabularx}
\usepackage{xltabular}

\renewcommand\thesubsubsection{\thesection.\arabic{subsection}.(\alph{subsubsection})}

\renewcommand\theparagraph{\thesection.\arabic{subsection}.(\alph{subsubsection}.\roman{paragraph})}

\usepackage{citeall}

\titleformat{\paragraph}[block] 
  {\normalfont\normalsize\bfseries}
  {\theparagraph}  
  {1em}  
  {} 
\titlespacing*{\paragraph}{0pt}{1ex}{0.5ex}

\titleformat{\subsubsection}[block]
  {\normalfont\normalsize\bfseries}  
  {\thesubsubsection}                
  {1em}                              
  {}     

\usepackage{hyperref}
\hypersetup{
colorlinks=true,citecolor=apolloblue,linkcolor=apolloblue,urlcolor=apolloblue
}

\title{The Loss of Control Playbook:\\ Degrees, Dynamics, and Preparedness}

\author{
Charlotte Stix \thanks{Correspondence to charlotte@apolloresearch.ai.} \and\hspace{-2em} Annika Hallensleben \and\hspace{-2em} Alejandro Ortega
\AND Matteo Pistillo 
\AND 
\textmd{Apollo Research}
}

\iclrfinalcopy 
\begin{document}
\vspace{2cm}
\maketitle

\input{content/Executive_Summary}

\input{content/Introduction}

\input{content/Chapter1}

\input{content/Chapter2}

\input{content/Chapter3}

\input{content/Conclusion}

\vspace{2em}
\subsection*{Acknowledgements}
For thoughtful feedback and valuable discussions related to this work, we would like to thank: David Africa, Jacob Arbeid, Benjamin Boudreaux, Axel Højmark, Marius Hobbhahn, Alex Meinke, Sören Mindermann, Jérémy  Scheurer, Anna Wang, Teun van der Weij, Caleb Withers, Mikita Balesni and multiple reviewers who wish to stay anonymous.

Reviewers mentioned may not agree with all claims made in this report. All errors remain our own.

\newpage
\appendix
\input{content/Appendix}

\begin{CJK*}{UTF8}{gbsn}
\printbibliography
\end{CJK*}

\end{document}

%% file: content/Executive_Summary.tex
\vspace{3cm}
\section*{Executive Summary}

This research report adopts a mission-oriented approach to conceptualizing and building preparedness for future loss of control (LoC) threats. It offers a practical analysis of the degrees and dynamics of LoC, as well as actionable tools that decision- and policymakers can leverage today to increase readiness for tomorrow’s national security and societal threats posed by LoC. 

LoC has become a topic of increased attention for policymakers on both sides of the Atlantic. Several recent legal frameworks, including California Senate Bill 53, the AI Risk Evaluation Act introduced by Senators Josh Hawley and Richard Blumenthal, and the EU AI Act’s General-Purpose AI Code of Practice, address LoC and include obligations for developers to assess and mitigate relevant threats.

\textbf{Despite the rising global attention, decision- and policymakers are still operating in the absence of an actionable definition of LoC}. This could, in turn, both encourage ‘crying wolf’ situations for scenarios that fall short of LoC, and prevent stakeholders from accurately forecasting and assessing early warning signs of LoC. In fact, existing LoC definitions vary on a wide range of parameters, resulting in even the two most consensus-based definitions––the definitions in the International AI Safety Report and the European Union AI Act’s Code of Practice for General-Purpose AI Models––to differ in both the spectrum of LoC outcomes covered, and the expected timelines for these outcomes.

\begin{figure}[h!]
   \centering
   \includegraphics[width=0.8\linewidth]{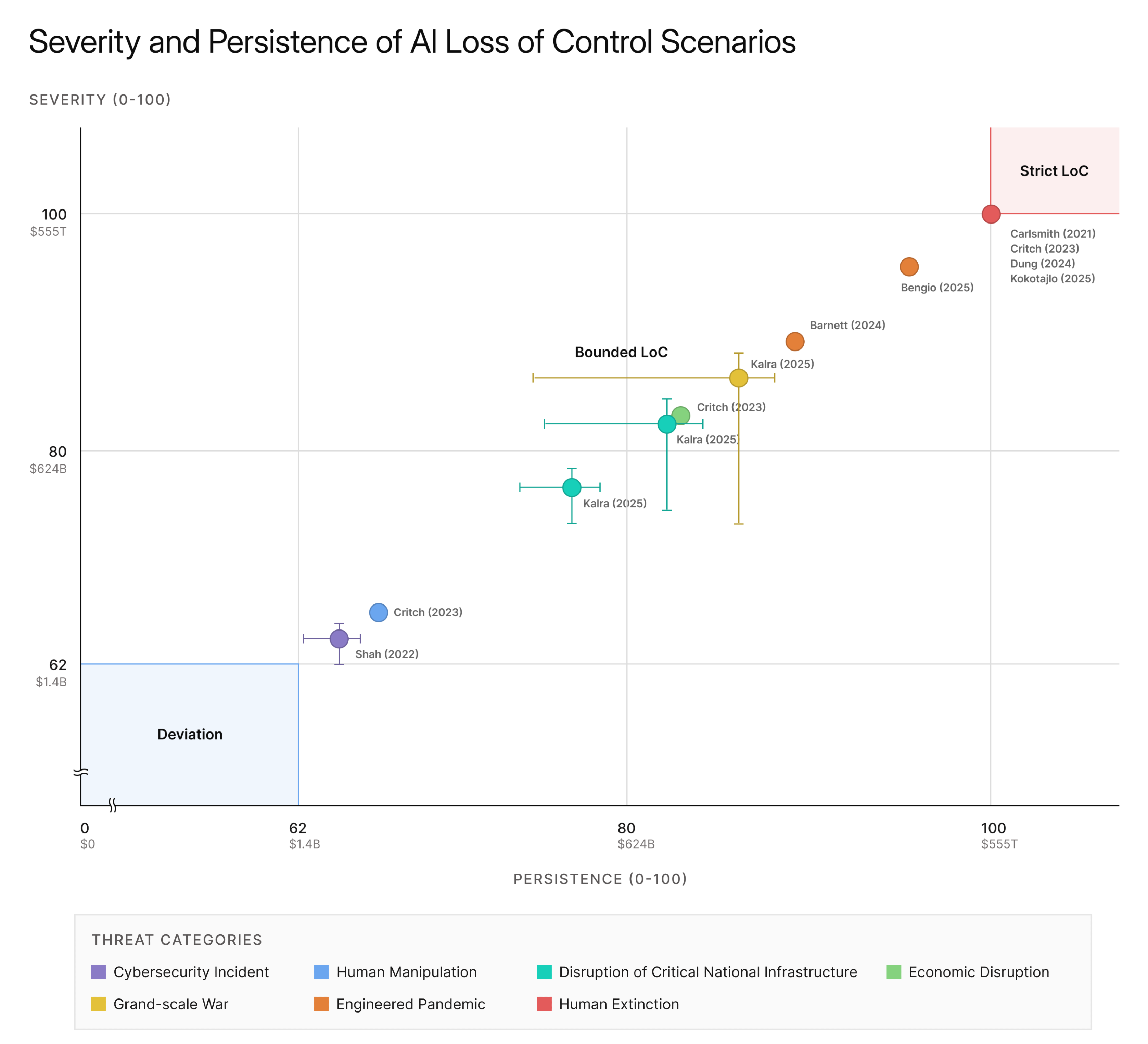}
   \caption{This graph plots 12 concrete LoC scenarios identified in the literature. We utilized economic impact as a proxy for severity and persistence, which are both mapped on arbitrary axes of 0-100.  These data points inform our proposed three-part taxonomy, splitting apart Deviation, Bounded LoC, and Strict LoC. By sharpening the conceptual boundaries of LoC,  this taxonomy helps decision-makers to understand the different degrees of  LoC and hence better prioritise between risk-reduction strategies. See \hyperref[findings_loc_lit]{Section 1.A.2} for more detail on the methodology behind and limitations of this visualization.}
   \label{figure_exec:the_graph}
\end{figure}
\addtocounter{figure}{-1}

\textbf{This research report aims to address existing divergences and make current LoC definitions and conceptualizations action-ready by demonstrating that it is possible to conceptualize multiple ‘degrees’ of LoC}. We present these degrees through a novel taxonomy of LoC based on an extensive literature review and methodology that allows us to place concrete LoC scenarios along two axes: severity (i.e., how many people are affected by a LoC event and to what degree they are affected), and persistence (i.e., how difficult it is to interrupt the ‘harm trajectory’ of a LoC event). Our resulting taxonomy helps decision- and policymakers visualize the lower and upper boundaries of the LoC spectrum more clearly, and better conceptualize the near- and longer-term threats that LoC could pose to national security and humanity. Specifically, drawing from our literature review, our taxonomy distinguishes between the following three degrees:
\newpage
\begin{itemize}
    \item \textbf{Deviation}: captures events that cause some harm or inconvenience but lack the requisite severity and persistence to reach the economic consequences threshold that the the U.S.'s Department of Homeland Security, Intelligence Community, and other components use to demarcate national-level events in the Strategic National Risk Assessment.
    \item \textbf{Bounded LoC}: captures events that cause great damage or suffering, and are difficult but not impossible to contain, albeit potentially at great cost. Bounded LOC captures threats or hazards that could have the potential to significantly impact the U.S. homeland security. 
    \item \textbf{Strict LoC}: captures events that are maximally severe and permanent, capturing events that result in humanity as a whole becoming extinct. 
\end{itemize}

We subsequently put forward a ‘playbook’ that can help decision- and policy-makers build \textbf{straightforward and actionable preparedness for LoC }today and in the near future. Given existing limitations in the current understanding of the role of capabilities and propensities (the ‘intrinsic’ factors) in contributing to LoC, we instead focus our attention on the LoC dynamics enabled by ‘extrinsic’ factors. Specifically, we focus on an AI system’s:\footnote{ We follow the definition of an AI system as described in \parencite{Sharkey2024d} capturing AI systems to “include not only the weights and architecture of the AI system, but also include a broader set of system
Parameters [...]. These consist of retrieval databases and particular kinds of prompts.”} (i) \textbf{deployment context}, meaning the combination of a given AI system's intended use case and the specific environment within which an AI system is deployed; (ii) \textbf{affordances}, meaning the environmental resources and opportunities for affecting the world available to an AI system; and (iii) \textbf{permissions}, meaning the set of authorizations an AI system is given to exercise its capabilities through the available affordances. We refer to this focus on leveraging extrinsic factors to manage LoC threats as a ‘DAP framework’. For each component, we suggest the following steps: 

\begin{itemize}
    \item \textbf{Deployment context}: (1) reviewing the ‘composition’ of the deployment context (i.e., the environment and use case), and clarifying whether the deployment context should be considered as ‘high-stakes’ (e.g., critical national infrastructure, military, AI research and development) or not; and (2) assessing the potential for cascading failures across interconnected AI- and non-AI systems, including through threat modeling and red teaming.

    \item \textbf{Affordances}: (1) considering whether an affordance is necessary to achieve the intended task; (2) considering every action that an affordance could enable, and the negative consequences thereof, and, consequently, limiting the affordance as much as feasible to reduce the risks through permissions; and (3) accounting for the potential for future, highly advanced AI systems to manipulate insufficiently informed human users into giving the AI system additional affordances.
    
    \item \textbf{Permissions}: (1) restricting permissions to the minimum necessary for an AI system to complete the task, taking into account the well-established principle of least privilege; (2) weighing the benefits and risks of a human’s reduced oversight against the benefits and risks increased permissions could bring; and (3) accounting for the potential for future, highly advanced AI systems to manipulate insufficiently informed human users into giving the AI system additional permissions.
\end{itemize}

We then reflect on a set of societal and technical dynamics and how these could affect future resilience to LoC. Specifically, we reflect on the likelihood of continuous AI capability progress, and on the growing economic and strategic pressures and incentives to leverage AI systems in more complex and high-stakes deployment contexts, endowing them with broader affordances and permissions. We propose that unless these dynamics are handled strategically, it is likely that \textbf{society will eventually encounter a }‘\textbf{state of vulnerability}.’ We use the expression ‘state of vulnerability’ to describe a necessary precondition in which future, highly advanced AI systems have acquired or could independently acquire sufficient access to resources, affordances, permissions, and capabilities to cause LoC once a catalyst materializes. As a LoC catalyst, we envision both: (i) malfunctions that are \textbf{misalignment}; and (ii) \textbf{malfunctions that are not misalignment} (‘pure malfunctions’). 

Subsequently, \textbf{we model several theoretical pathways that a future with a state of vulnerability could take with regard to the threat of LoC}. In doing so, we propose that it is highly unlikely society would not eventually encounter a LoC threat. As part of our theoretical model, we suggest that it is unlikely that the core catalyst of LoC (misalignment or pure malfunctions) would be resolved. In fact, we propose that, even if the alignment problem were solved, pure malfunctions might still occasionally occur, since it is difficult to ensure their non-occurrence ex ante, which makes a future where society lives with a state of vulnerability precarious. Finally, we suggest that if humanity reaches a state of vulnerability, preparedness means being able to \textbf{hold the state of vulnerability in a condition of perennial suspension}, including through a combination of:

\begin{itemize}
    \item \textbf{Governance interventions}, such as: (1) concrete threat modelling; (2) policies describing acceptable deployments; and (3) wide-reaching, easy-to-enact emergency response plans.
    \item \textbf{Technical interventions}, such as: (1) rigorous pre-deployment testing suites in accordance with threat models for the deployment context; (2) control measures that constrain an AI system’s effect on the world around the AI system; and (3) stringent human and AI-enabled monitoring. 
\end{itemize}

%% file: content/Introduction.tex
\phantomsection
\section*{Introduction}\label{introduction}

In this research report, we provide novel perspectives on what we perceive to be one of the most challenging topics in the field of technical AI governance: loss of control (LoC). Our work is motivated by recent growing attention to LoC, as well as the expectation that, in light of continued AI progress, the landscape for LoC threats will soon become more critical, necessitating greater specificity and conceptual clarity than is currently available. 

LoC has become a topic of increased attention for policymakers on both sides of the Atlantic. In the United States (U.S.), California enacted Senate Bill 53, which includes LoC within the “critical safety incidents” that frontier AI developers should identify, respond to, and report \parencite{CASB53}, reflecting a previous mention of LoC in the California Report on Frontier AI Policy \parencite{CA_Frontier_AI_Policy_2025}. More recently, the bipartisan legislation introduced by Senators Hawley and Blumenthal identifies LoC scenarios as one of the adverse AI incidents whose likelihood the Department of Energy would be tasked to evaluate \parencite{AIREA}. In the European Union (EU), the EU AI Act’s General-Purpose AI Code of Practice categorizes LoC as a systemic risk, requiring covered deployers to assess and mitigate its risk (\cite{EU_AI_COP}, Safety and Security Chapter, Appendix 1.4).

Similarly, LoC has been the subject of increasing attention from various frontier AI companies. Some of these companies have committed to voluntarily assessing AI capabilities instrumental to AI systems’ potential to “undermine human control” \parencite{FSF_GoogleDeepMind2025} in their Frontier Safety Policies, and some have included LoC within the societal impacts discussed in their System Cards \parencite{GPT-4o_OpenAISystemCard}.

Notwithstanding the rising interest, the topic area of LoC itself is nascent and lacks a common, clear definition. In this context, we found that neither AI literature nor other safety-critical industries (\textit{see} \hyperref[findings_loc_def]{Section 1.A.1}) provide sufficient support for a common, clear, and, in turn, widely actionable definition of LoC. The absence of such a definition may soon contribute to epistemic confusion around LoC, especially as more decision- and policymakers engage with the topic. In turn, this could make the identification of early warning shots challenging and undermine the shared effort by the aforementioned and future legal frameworks to target and mitigate LoC threats appropriately. Finally, an absence of clarity could also lead to ‘crying wolf’ situations for phenomena that fall short of LoC, and ultimately contribute to fostering skepticism of LoC itself as a remote, impalpable, and speculative threat. 

It is plausible that the object-level threat of LoC will significantly grow in the coming couple of years. We expect that this will occur due to two reasons: (i) AI capabilities will likely increase in line with established trends \parencite{epoch2025gptcapabilitiesprogress,Task_Horizon_METR,epoch_prediction}; and, concurrently, (ii) there will be growing economic and strategic incentives to use these more highly advanced future AI systems in areas that are currently too complex or costly for wide AI system deployment \parencite{GDP_Eval_OpenAI,Patel2024k, shah2025approachtechnicalagisafety}. As a result, it is likely that these AI systems will be integrated into broader and higher-stakes use cases––some of which could be critical, such as in certain defense applications or in sensitive government settings \parencite{CDAO_Partnerships_FrontierAI_2025, Anthropic_Government_Solutions,UK_MOD_AI_Trial_2025}––and be given more wide-reaching modalities to execute on their use cases. Ultimately, this combination of advanced capabilities and AI system integration into high-stakes use cases in critical sectors could lead to more significant, far-reaching impacts should an undesirable event occur, including those that can devolve into LoC. 

In this research report, we attempt to bring all aforementioned challenges into coherence. In doing so, we offer:  \begin{itemize} [itemsep=0.6pt,parsep=0.6pt]
    \item \textbf{A novel taxonomy of LoC} to interpret existing definitions and categorize examples from AI literature, as well as to pinpoint the levels of severity and persistence most commonly referred to in scholarly discussions of LoC threats (\textit{see} \hyperref[chapter-1]{Chapter 1}).
    \item \textbf{A straightforward framework to minimize the threat of LoC today}, which works around existing uncertainties regarding LoC-relevant capabilities and propensities, and that decision- and policymakers can, therefore, leverage immediately (\textit{see} \hyperref{chapter-2}{Chapter 2}). 
    \item \textbf{A theoretical argument as to how likely it is that society will eventually encounter a LoC threat}, based on the plausibility of arriving at a societal ‘state of vulnerability’ to LoC and its implications for our future (see Chapter 3).
\end{itemize}

We endeavored to adhere to the core goal of this research report and capture the limited but growing existing evidence on LoC, without engaging in broad speculation. The scope of this research report is, therefore, limited as follows:
\begin{itemize} [itemsep=0.6pt,parsep=0.6pt]
    \item We do not provide a quantitative estimate of risk or an assessment of the likelihood of specific LoC scenarios described in the broader literature. 
    \item We do not elaborate on how likely it is that a specific capability highlighted in recent research \parencite{phuong2025evaluatingfrontiermodelsstealth, black2025replibench, meinke2025incontextscheming} would lead to LoC, nor on capability composition or thresholds.
\end{itemize}

Finally, throughout our research, we came across a variety of important open questions that we elected not to discuss and situate in this report. We offer a list of these questions in \hyperref[appendix1]{Appendix 1}.

%% file: content/Chapter1.tex
\phantomsection
\section*{Chapter 1. A Taxonomy of Loss of Control}\label{chapter-1}
\captionsetup{justification=justified}
In this Chapter, \textbf{we put forward a novel taxonomy of LoC}. 

In order for us to arrive at this taxonomy, we first reviewed a range of existing definitions of LoC across the broader AI literature. Our goal from this initial review was to understand whether there exists sufficient consensus within AI literature around a definition of LoC that could be operationalized by decision- and policymakers today. As we will explain below, we did not find satisfactory results in that direction. Instead, we found that existing definitions of LoC in literature differ across a range of axes and could be interpreted to cover different spectra of outcomes. We subsequently examined whether LoC definitions and terminology in other safety-critical sectors, such as defense, aviation, and nuclear, could transfer to AI and provide additional interpretative lenses. As this review also failed to provide definitive answers on an actionable definition of LoC for AI, we decided to home in on definitions characterized by their common element of having been shaped by large-scale multi-stakeholder consultations and consensus-driven processes—the EU AI Act’s General-Purpose AI Code of Practice (COP, \parencite{EU_AI_COP}) and the International AI Safety Report (IASR, 
\parencite{IASR_2025}). After observing significant differences between the definitions espoused in these two publications, we turned towards a comprehensive review of literature relevant to LoC, alongside an assessment of the scenarios contained therein.

Concretely, based on this process and its outcomes, we derived a novel conceptualization of LoC and inferred that there are at least three broad categories. We taxonomize these three as Deviation, Bounded LoC, and Strict LoC. They are conceptualized as follows:\footnote{ We note that none of the boundaries proposed in our taxonomy are supposed to be construed as exact. They are intended to provide intuitions to talk about LoC more adequately.}

\begin{itemize}[itemsep=0.6pt,parsep=0.6pt]
    \item \textbf{Deviation} captures events that cause some harm or inconvenience, but which are relatively easy to contain.
    \item\textbf{Bounded LoC} captures events that can cause great damage or suffering, and are difficult, but possible, to contain, albeit potentially at great cost.
    \item \textbf{Strict LoC} captures events that are maximally severe and permanent, such as events that result in humanity as a whole becoming extinct.
\end{itemize}

The remainder of Chapter 1 will describe our overall process, methodology, and how we arrived at this taxonomy and its implications in more detail. 

\phantomsection
\subsection*{1.A.1 Findings from Loss of Control Definitions in Literature}\label{findings_loc_def}

A range of governance frameworks and research papers have sought to define LoC in the context of highly advanced future AI systems. Yet, there exists no overarching consensus on the precise meaning of LoC, as we shall demonstrate, posing a challenge for decision- and policymakers alike. We briefly present a selection of definitions from the literature and describe several key learnings from our review, before delving into more depth on two definitions in particular.

In general, LoC definitions refer to situations in which humans lose the ability to effectively manage, direct, or intervene in the operation of increasingly capable AI systems. The following is a range of non-exhaustive examples:

\begin{itemize}
    \item The Singapore Consensus on Global AI Safety Research Priorities defines LoC as “... scenarios where advanced AI systems – such as AGI – come to operate outside of human control, with no clear path to regaining control” \parencite{bengio2025singaporeconsensusglobalai}.
    \item The Gladstone Action Plan defines LoC as a “failure mode under which a future AI system could become so capable that it escapes all human efforts to contain its impact” \parencite{Gladstone_ActionPlan_2024}.
    \item An emergency preparedness report by RAND defines LoC as “situations where human oversight fails to adequately constrain an autonomous, general-purpose AI, leading to unintended and potentially catastrophic consequences” \parencite{RAND_Preparedness}.
    \item Legislation recently introduced by Senators Hawley and Blumenthal defines LoC as “a scenario in which an artificial intelligence system: behaves contrary to its instruction or programming by human designers or operators; deviates from rules established by human designers or operators; alters operational rules or safety constraints without authorization; operates beyond the scope intended by human designers or operators; pursues goals that are different from those intended by human designers or operators; subverts oversight or shutdown mechanisms; or otherwise behaves in an unpredictable manner so as to be harmful to humanity” \parencite{AIREA}.
    \item The consensus statement signed by global experts from academia, AI companies, and independent organizations as a result of the 2025 International Dialogues on AI Safety (IDAIS) in Shanghai describes loss of control as a situation in which “one or more general-purpose AI systems come to operate outside of anyone’s control, posing catastrophic and existential risks” \parencite{IDAIS_Shanghai_2025}.
\end{itemize}

\textbf{Despite certain commonalities, such as concerns about AI systems operating beyond the scope of reliable human direction or oversight, definitions in the broader AI literature differ in their emphasis and framing} \parencite{bernardi2025societaladaptationadvancedai, kulveit2025gradualdisempowermentsystemicexistential, UK_DSIT_FrontierAI_2023, CIGI_Framework_Convention_2024}\footnote{Due to our methodology, as described later in this Chapter, human disempowerment, due to its uncertain nature, is out of the scope of this report.} \textbf{thereby presenting conceptual distinctions.} Divergences between definitions can make it difficult to implement functional frameworks and enact suitable interventions. One axis of variation relates to the cognitive capabilities required by an AI system for LoC. Some researchers appear to suggest LoC may occur once AI systems become more intelligent than humans \parencite{hendrycks2023overviewcatastrophicairisks, MIRI_Mission_Strategy_Update_2024}. Others more explicitly refer to artificial superintelligence when describing LoC risk \parencite{RAND_Rise_Fall_Nations, barnett2025aigovernanceavoidextinction}. In contrast, Stuart Russell’s definition in ‘Artificial Intelligence and the Problem of Control’ requires only a “sufficiently capable machine” \parencite{Russell2022}, rather than one explicitly surpassing human intelligence or achieving superintelligence. A further divergence across definitions involves the difference in treatment between oversight and control. While some definitions refer to control \parencite{CA_Frontier_AI_Policy_2025, IDAIS_Shanghai_2025}, others emphasize oversight instead, focusing on AI systems that “operat[e] beyond human oversight” \parencite{METR_AGI_Definitions_2025}. Another axis of variation concerns the possibility of regaining control. For example, the COP \parencite{EU_AI_COP} does not necessarily imply that LoC is irreversible, whereas the IASR \parencite{IASR_2025}, the Singapore Consensus \parencite{bengio2025singaporeconsensusglobalai}, and other reports \parencite{Bengio_managing_risks_2024, hendrycks2023overviewcatastrophicairisks} all suggest that LoC involves the absence of a clear path to regaining control. These interpretations imply that once control is lost, recovery may be impossible or extremely difficult. Overall, we found that while definitions aim to cover similar areas, they do so in sufficiently diverse ways to prevent a common, clear definition of LoC arising from the AI literature.

\textbf{We observe that the lack of consensus surrounding one concrete definition of LoC matches what we learned from reviewing the concept of LoC in other safety-critical sectors. }Specifically, we found that the meaning of LoC is not unified across different sectors, and spans from unauthorized access to personally identifiable information \parencite{OMB_M_17_12,NIST_breach_definition, CISA_Playbooks_2021}  in the cybersecurity sector, to a yaw motion leading to a deviation from a driver’s intended path \parencite{FMVSS_136_CFR_2015}  in the automotive sector, to the uncontrollable diffusion of chronic/latent pathogen infections \parencite{FDA-Immunotoxicity-2020-Draft} in the pharmaceutical sector.

In the majority of cases, \textbf{LoC takes on different meanings even within the same sector, depending on the circumstances}. For instance, in the defense sector, the term LoC can refer to adversaries gaining control of an autonomous or nuclear weapon \parencite{DoD3000_09, DoDD3150.08,DoDM-S-5210-41-Vol1-2016, DoD-CWMD-Strategy-2014,DoD-QDR-2006}, to information being lost, stolen, or compromised \parencite{DoD5400.11R}, to mishaps and near-mishaps encountered during air combat \parencite{OPNAVINST-3750-6R-CH4-2009}, and to interference with command and control \parencite{DoDDictionary2021,NIST-SP-800-59-2003}. In aviation and space, LoC can refer to an aircraft’s deviation from the intended flightpath \parencite{FAA-FlySafe-LOC-2019, CICTT_AviationOccurrenceCategories_2013, FAA_AC_120_111_CHG1, EASA_LOC_I, CAST_JSAT_LOC_2000, FAA_SAFO_17009, FAA_Handbook_Ch5}, and to a remote pilot not being able to command an unmanned aircraft \parencite{CFR14-107-19, sakakeeny_lost_command, hayashi2022paav, sakakeeny2022framework}, but also to an uncontrollable space station following collision with debris or meteoroids \parencite{CFR47-97-207-2025a,CFR47-5-64_satellite_systems,CFR47-25-114_space_auth}. In nuclear, LoC can refer to nuclear chain reactions leading to an explosion such as the  Chernobyl Nuclear Power Plant (\cite{IAEA_Chornobyl_Topic}; Requirement 16, \cite{IAEA_SF1_2006}; \cite{NRC_Bulletin_91_01}), or can also refer to the loss of licensed radioactive sources, such as nuclear fuel (\cite{CFR10-20-2202b_notification_incidents}; \cite{NUREG_1794_2004}; \cite{NRC_Koch_Civil_Penalty_1998}; \cite{10-CFR-Chapter-I}; Requirement 80,  \cite{IAEA_SF1_2006}; \cite{ortiz2002lessons}; \cite{international2004strengthening}). The divergent and broad nature of definitions of LoC across and within other safety-critical sectors limits their usefulness for the AI sector. The plausible future nature of AI as an agentic system ‘driving’ LoC presents a particular nuance that other sectors have not yet meaningfully accounted for.

In order to have some anchor point for our work, we ultimately chose to home in on the definition in the COP \parencite{EU_AI_COP}  and the definition in the IASR \parencite{IASR_2025}. We focus on these two definitions because they have been developed by a broad range of technical and policy stakeholders, undergoing multiple iterations and consensus-building, which speaks to the inclusion of diverse views already within each definition. In the case of the COP, the LoC definition also underpins concrete regulatory interventions.

\begin{itemize}
    \item The EU AI Act’s Code of Practice for General-Purpose AI Models defines LoC as “risks from humans losing the ability to reliably direct, modify, or shut down a model” (COP, \cite{EU_AI_COP}).
    \item The International AI Safety Report defines LoC as “...scenarios in which one or more general-purpose AI systems come to operate outside of anyone’s control, with no clear path to regaining control” (IASR, \cite{IASR_2025}).
\end{itemize}

In reviewing these two definitions specifically, we make two observations. First, we note that both definitions can be interpreted to imply a\textbf{ spectrum of outcomes across both scale and severity}. For instance, the COP definition can be interpreted to capture both more or less consequential outcomes, all arising from an inability to reliably affect or shut down an AI system. Similarly, across a spectrum of severity and persistence, under the IASR definition, LoC occurs if humanity has “no clear path to regaining control,” which leaves open the question whether a scenario in which regaining control is extremely difficult and costly (but still possible) falls within the LoC definition.

Second, we note that the two definitions\textbf{ differ in the expected timelines }for the LoC outcomes they aim to capture implicitly. For instance, under the COP definition, LoC materializes when humans lose the ability to “reliably direct” an AI system. Arguably, based on this definition, we can already see instances of LoC occurring today. For instance, in a situation where an AI system cheats to obtain artificially high scores on a test, one could argue that the user is unable to reliably direct the AI system to complete the task \parencite{Reward_Hacking_METR}. These occurrences are no longer limited to testing environments; for example, a coding AI system recently wiped an entire production database, despite being explicitly instructed not to make any changes \parencite{Cybernews_Replit_VibeCoding_2025}. While such an outcome would be captured under the COP definition, it would not be captured by the IASR definition. Specifically, by contrast, the IASR definition seems to require that there must be “no clear path to regaining control” for a scenario to constitute LoC. Therefore, the definition captures only scenarios in which regaining control is, if not impossible, at least, by way of interpretation, very difficult and costly. Our previous example would not be covered by this definition, even using its broadest interpretation, since regaining control was simple despite the AI system not acting reliably. In other words, it seems implausible that AI systems that are already on the market would be captured by the IASR definition, whereas this does not seem implausible for the COP definition. We therefore conclude that, despite having undergone broad consensus-building processes, both definitions are at odds across certain dimensions.\footnote{This is not to be misconstrued as an assessment of adequacy for either of these definitions. Such an assessment is out of the scope of this report.}

The aforementioned challenges and limitations surrounding the definition of LoC for AI systems prompted us to review existing AI literature describing LoC scenarios and their outcomes. The goal of that review (\textit{see} \hyperref[findings_loc_lit]{Section 1.A.2}) was to produce a more comprehensive and nuanced picture of what AI researchers mean when they discuss LoC non-abstractly. By extension, we expected that this additional review could serve to supplement leading multi-stakeholder-developed definitions, such as those in the COP or IASR, enable a more lucid dialogue for decision-makers, and allow for more specification to enable actionable measures against what we eventually found to be different categories of LoC. 
\phantomsection
\subsection*{1.A.2 Findings from Loss of Control Scenarios in Literature}\label{findings_loc_lit}

Our literature review was motivated by the goal of adding nuance and clarity to the concept of LoC, and by doing so, assessing which outcomes the field considers to be LoC and whether there are any overarching and notable commonalities that can further advance the conceptualization of LoC beyond existing efforts. We describe our methodology and process first, and then we present our findings.
\phantomsection
\subsubsection*{1.A.2.a Methodology}\label{1A2a_methodology}
In total, we reviewed 130 works from academia, international think tanks, and government agencies (\textit{see} \hyperref[appendix2.1]{Appendix 2.1} for the full list).\footnote{These works were selected by filtering for those that reflect on implications of future AI progress, covering diverse threats, risk assessments, and capabilities commonly described as control-undermining (\textit{see} \hyperref[2A_capa_prop]{Section 2.A}). Some works were further selected because they appeared in relevant passages in the literature we were reviewing.} We subsequently applied three filters to these works in order to arrive at comparable and informative data points: (i) do they contain a scenario; (ii) does the scenario concern LoC; (iii) is the LoC outcome sufficiently concrete for us to derive learnings from. We describe each step in more detail next. 

First, \textbf{we filtered works into two categories}: \textbf{those that contained scenarios and those that did not}. In order to count as a scenario, a text passage within a given piece of literature had to pass our causal detail criterion. In other words, a text passage was considered a scenario if it could fulfill the causal detail criterion by either: (i) containing a detailed narrative description of the events leading up to the outcome; or (ii) giving an abstract but highly detailed logical argument about how an AI system could cause a certain outcome.

\textbf{Our next filter assessed whether the scenarios could be reasonably considered LoC scenarios}, as most reviewed works did not, in fact, mention this keyword.\footnote{We suspect that the reason for this is that LoC is a relatively new term to cover a specific threat category and, therefore, only appears in very recent literature.} We assessed the scenarios against four definitions across four governance documents, previously described in \hyperref[findings_loc_def]{Section 1.A.1}. Specifically, we assessed them against the definitions in the AI Act's Code of Practice for General-Purpose AI Models \parencite{EU_AI_COP}, the Artificial Intelligence Risk Evaluation Act of 2025 \parencite{AIREA}, the International AI Safety Report \parencite{IASR_2025}, and the Singapore Consensus on Global AI Safety Research Priorities \parencite{bengio2025singaporeconsensusglobalai}. In selecting these four definitions, we aimed to achieve a balance of definitions that are regulatory in nature and definitions that have directly benefited from a broad-ranging, international, and multidisciplinary perspective, developed by contributors from across industry, academia, and government. If a scenario could reasonably be captured by any of these four definitions, it passed our LoC filter. This left us with 40 individual LoC\footnote{ In our report, a LoC scenario captures a negative outcome resulting from LoC. Positive types of outcomes are therefore outside of the report’s scope.} scenarios in total. 

Our \textbf{final filter for these LoC scenarios was to assess whether we could deem them sufficiently concrete to allow for meaningful comparison between scenarios.} In order to find a reasonable and overarching filter for assessing sufficient concreteness, we drew on two dimensions we identified as present across all LoC scenarios. We conceptualize these two metrics as follows: (1) \textbf{severity, capturing how many people are affected and the degree to which they are affected}; and (2) \textbf{persistence, capturing the difficulty of interrupting the ‘harm trajectory’ of the scenario}.\footnote{To use a metaphor, interrupting the harm trajectory would mean interrupting a sequence of falling dominos mid-cascade, but not standing the fallen ones back up.} More concretely, we conceptualize persistence as consisting of two components: (i) the difficulty of preventing the AI system from taking additional unintended actions that cause further harm; and (ii) the difficulty of interrupting the immediate harmful process initiated by the AI system. 

We propose that both metrics—severity and persistence—can be captured by economic impact as a proxy measure and expand on our reasoning in more detail in the subsequent paragraphs. 

\textbf{For severity, we propose that economic impact can serve as a suitable proxy because the more people are affected and the more severely they are affected, the more plausible it is that the disaster will have had a larger economic impact}. Various works in academic literature offer evidence to support our claim that severity can be expressed by economic impact, for example, in events such as power-grid interruptions \parencite{larsen2025ice}, pandemics \parencite{konig2021covid,morgenstern2024interaction}, wars \parencite{mueller2016cost, NovtaPugacheva2020}, and natural disasters \parencite{cavallo2022impact}, including, for example,  hurricanes \parencite{hsiang2014causal, acevedo2016gone, murnane2012maximum, Huang2024TropicalCycloneMortality}. Similarly, adverse effects of climate change \parencite{NORDHAUS20131069, estimating_econ} and health impacts, such as the burden of pollution-related health impacts, are often expressed in economic terms \parencite{deryugina2019mortality, chang2016particulate}. As these events escalate, their economic impacts also escalate.

\textbf{For persistence, we propose that economic impact is a suitable proxy because it is plausible that more persistent scenarios take longer to interrupt, and, in turn, scenarios that take longer to interrupt can plausibly be considered to have a larger economic impact.} Scenarios that are more persistent have harm trajectories that are more difficult to interrupt. In other words, it is more challenging to interrupt the AI system itself, which causes the scenario, and the immediately harmful process that results from the AI system's actions. Persistence across the literature appears to be correlated with the relative ease or difficulty in resolving hindrances to interrupting the harm trajectory, such as an absence of appropriate technical knowledge, other resource constraints, or coordination failures. Due to their nature, these hindrances generally present time-consuming hurdles, disallowing a swift interruption of the harm trajectory. We propose that it is reasonable to assume that scenarios where the harm trajectory takes longer to interrupt have a greater economic impact, as the occurrence of harm spans a longer time horizon. Various works in academic literature offer evidence to support our claim, indicating a causal link between the duration of an incident and the economic impact it causes. For example, in electricity-grid management there is some evidence that the economic impact of electricity blackouts rises with the duration of the blackout \parencite{ENWL-VoLL-Phase3-2018,LBNL-6941E-2015}; there is evidence that longer duration heatwaves \parencite{OECD-Heat-Stress-2024} are associated with a higher economic impact, and literature suggests that longer-lasting recessions \parencite{haltmaier2013recessions} and bank crises have larger impacts on the economy \parencite{hoggarth2001costs}. 

Drawing on our conceptualization and contextualization of severity and persistence led us to \textbf{adopt economic impact, which is standardized and readily available, as a proxy measure and concreteness criterion to filter out concrete LoC scenarios}.

Therefore, in order to be deemed \textbf{concrete, a LoC scenario had to contain sufficient detail of the scenario’s LoC outcome such that we could estimate the economic impact of its outcome.}\footnote{ We focused on the outcome because an overwhelming number of scenarios have multiple, often cascading, mechanisms that were largely impossible to consistently and clearly delineate, categorize or estimate.} Any given LoC scenario could achieve the concreteness criterion via two methods: (i) if we were able to match it to a pre-existing economic estimate for the same or a highly similar scenario in literature––for example, derived from academic literature or third-party research reports (e.g., from \cite{RHG_Taiwan_Disruptions_2022} and \cite{posner2004catastrophe}); or, alternatively (ii) if we were able to make our own back-of-the-envelope calculation (BOTEC) of the economic impact.\footnote{Note that half of our scenarios ended up being captured by (i) and the other half by (ii).} For these BOTECs we made appropriate assumptions about a given scenario (about, for instance, the scale or location of the outcome) that either (ii.a) allowed us to match the scenario to a same or highly similar scenario with pre-existing economic impact estimates; or, (ii.b) enabled us to leverage existing calculations to derive economic impact estimates where no corresponding example was found in literature (\textit{see} \hyperref[appendix2.2]{Appendix 2.2} and \hyperref[appendix2.2.1]{2.2.1}).\footnote{For some LoC scenarios, we encountered sufficiently large uncertainties that we considered it unreasonable to attempt to obtain a point estimate of the economic impact, as it would likely not have been meaningful.}

\textbf{In total, we found 12 text passages that fulfilled all of our criteria and were therefore classified as concrete LoC scenarios}.

\textbf{Once we completed this categorization, we filed the 12 concrete LoC scenarios within their respective ‘threat category.’} We use threat categories as a classification that groups LoC scenarios based on the outcome of the scenario, rather than narrative contributors to a given scenario. We found that the 12 concrete scenarios fell within the following threat categories:

\begin{itemize}[nosep]
    \item disruption of critical national infrastructure (CNI; \cite{ppd21_2013}; \cite{NPSA_CNI_2025}) \\
    (2 scenarios); 
    \item engineered pandemics (2 scenarios);
    \item grand-scale conflict/war (1 scenario);
    \item cybersecurity incident (1 scenario);
    \item economic disruption (1 scenario); 
    \item human manipulation (1 scenario); and,
    \item human extinction (4 scenarios).\footnote{We note that there may be a slight distortionary effect regarding human extinction as a threat category. Specifically, it is, by definition, very clear-cut and straightforward to calculate, making it significantly less challenging than other examples and, therefore, more concrete overall. It is plausible that the occurrence immediately preceding extinction should be captured, rather than the penultimate outcome, for these scenarios. While this would not align with our methodology and is therefore outside the scope of this report, it may be valuable for future scholarly research to consider.} 
\end{itemize}

\begin{figure}[h!]
   \centering
   \includegraphics[width=0.8\linewidth]{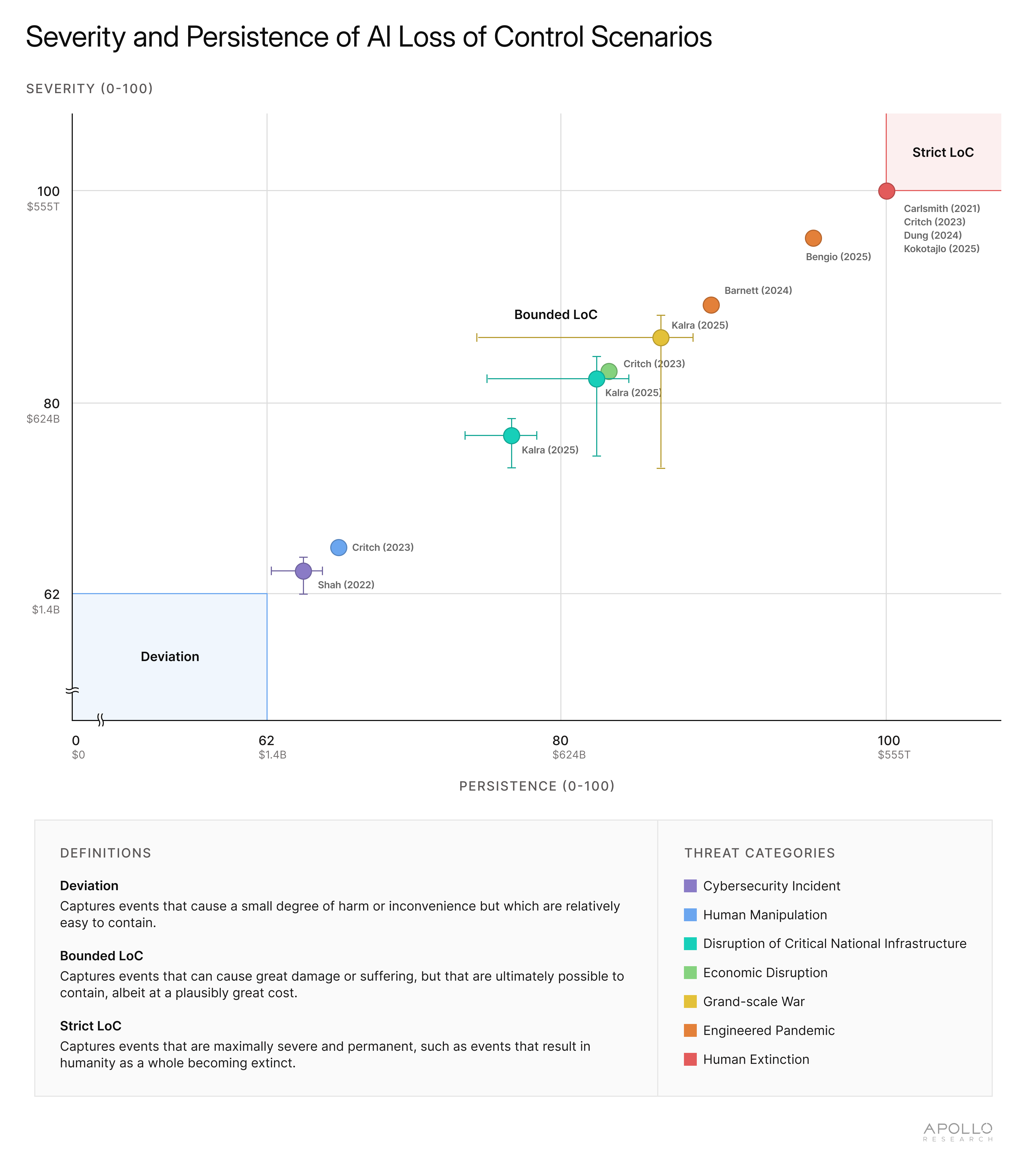}
   \caption{The distribution in this graph covers all 12 concrete LoC scenarios derived from the literature, plotted by severity and persistence using economic impact as a proxy measure (both axes in arbitrary units 0-100).\protect\footnotemark \space  The colors of scenarios indicate the relevant threat category: human extinction, human manipulation, economic disruption, cybersecurity incident, grand-scale conflict/war, engineered pandemic, or disruption of critical national infrastructure. Where multiple economic impact estimates for a scenario were available, error bars represent 50\% confidence intervals calculated using the t-distribution. For several scenarios, these error bars are too small to be visible on this log graph. For two scenarios, there are no error bars because only one estimate was available.}
   \label{figure2:the_graph}
\end{figure}

Next, \textbf{we plotted all 12 concrete LoC scenarios} on an experimental graph, using severity and persistence as the axes,\footnote{Given the difficulty of computing the precise economic impact for persistence, especially for concrete LoC scenarios from the literature that do not tend to consider this aspect in any detail, we decided to use the same economic impact estimate for persistence as for severity. We note that, while severity and persistence are not necessarily equivalent in terms of economic impact in all instances, we believe that they are quasi-equivalent for the scenarios plotted on our graph (\textit{see}~\autoref{figure2:the_graph}).
} and color-coded them within their respective threat categories (\textit{see}~\autoref{figure2:the_graph}). In doing so, we noted that \textbf{certain areas emerged from our plots and sought to contextualize them} within concepts and frameworks that already exist in governance and which decision- and policymakers may be familiar with.

\footnotetext{These arbitrary units were defined by setting the arbitrary unit 100-mark at \$550 trillion, and the 0-mark at \$1, and then creating a linear mapping between these two points in log-space. The linear mapping is performed in log space because the plotted scenarios vary in economic impact by four orders of magnitude, between \$30 billion and \$550 trillion. To convert from “Log Dollars” to “Arbitrary Units of Severity/Persistence,” we used the formula Arbitrary\_Severity (or Arbitrary\_Persistence) = $100 \times \log_{10}(\text{dollars}) / \log_{10}(5.5 \times 10^{14})$).} 

First, we sought to \textbf{contextualize the area indicated in the lower left }of our graph (\textit{see}~\autoref{figure2:the_graph}). In doing so, we came across the economic consequence thresholds that the U.S.'s Department of Homeland Security (DHS), the Intelligence Community, and other components, identified as “necessary to create a national level-event” in the Strategic National Risk Assessment (SNRA) \parencite{DHS_SNRA_2011}. Specifically, in 2011, DHS led an effort to “identify the types of incidents that pose the greatest threat to the Nation’s homeland security,” including various natural, technological, and adversarial, human-caused hazards \parencite{DHS_SNRA_2011}.\footnote{We decided to adopt the SNRA as a term of reference because the U.S. is the default location where the majority of literature assumes LoC scenarios to occur. For instance, ‘Scenario 1’ in \parencite{Kalra_GeopoliticsAGI} is an example of a scenario that envisages AI-caused electricity blackouts starting in the U.S. In some cases, the location of the LoC scenario described by literature is ambiguous (for instance, ‘Scenario 4’ in \parencite{Kalra_GeopoliticsAGI}), so for the purposes of making a comparable economic impact estimate, we assumed that the scenario occurred within the U.S. to ensure consistency across scenarios.} For example, earthquakes, floods, hurricanes, wildfires, and cyberattacks against physical infrastructure meet the threshold if they result in direct economic losses exceeding \$100 million \parencite{DHS_SNRA_2011}. Similarly, a cyberattack against data\footnote{ In this context, data refers to the information contained in a computer system or data processes.} is considered a national-level event if it results in economic losses of \$1 billion or more. In this research report, we adopt this last threshold set by the SNRA at \$1 billion or greater (adjusting it for inflation to approximately \$1.4 billion) to provide some contextualization of LoC scenarios (\textit{see}~\autoref{figure2:the_graph}). We refer to this threshold, i.e., the \$1.4 billion, for convenience, as the ‘\textbf{national risk assessment threshold}.’ In summary, the national risk assessment threshold is the highest threshold based on economic consequences that DHS, the Intelligence Community and other components 
identified in the Strategic National Risk Assessment to demarcate a national-level event (i.e., a threat or hazard that has the potential to significantly impact the U.S. homeland security). We indicate this threshold in blue in Figure 1.

Second, we sought to\textbf{ contextualize the area} \textbf{indicated in the top right} of our graph (\textit{see}~\autoref{figure2:the_graph}). In doing so, we drew on existing conceptualizations describing events of absolute scale and permanence. In other words, events that could lead to the destruction of humanity’s long-term potential \parencite{ord2020precipice, sundaram_mani_2025, stauffer2023existential, DallasFed_AI_Productivity_2025}.\footnote{ We note that we were unable to find a numerical operationalization for ‘existential catastrophe’ in the literature. Therefore, the boundary given on this graph is for illustrative purposes only, but is likely accurate, as it captures an absolute scale and permanence akin to the best interpretation of an existential catastrophe (\textit{see}~\autoref{figure2:the_graph}).} These events are commonly referred to as an \textbf{‘existential catastrophe,’} and we adopted this term to establish our upper boundary. In summary, existential catastrophe demarcates the point at which humanity loses control over its future in an absolute sense. We indicate this threshold in bold red dashes in Figure 1.

Reflecting on our methodology and choices, we note that while economic impact was the most tractable proxy for making calculations based on the LoC scenarios in the literature, it may not be the perfect proxy. We invite future scholarly research to establish a more refined methodology and calculations.\footnote{ We believe it is plausible that alternative methods exist for calculating these axes. For instance, severity could be estimated using a methodology wherein each affected person is assigned a number on a sliding scale between 0 and 1 to reflect the degree to which their health has been affected as a result of the scenario \parencite{SMC_QALY_Guide}. Alternatively, measurements could be based on wellbeing \parencite{frijters2024wellby}. These approaches would require significant further research to adapt their methods such that they could be used for AI LoC scenarios. This additional research was out of scope for this report. 
} We summarize our learnings next.
\phantomsection
\subsubsection*{1.A.2.b Reflections}\label{reflections_findings}

The aforementioned methodology allowed us to visually locate concrete scenarios derived from our literature review. The final plots helped identify the locus of attention that scholars have paid to LoC in literature, as well as its relation to existing conceptualizations of risk, such as existential catastrophe or a national risk assessment threshold.

Through the analysis enabled by this methodology, we found that: 

\begin{itemize}
    \item There exists a \textbf{comparatively low number of concrete LoC scenarios in literature}, vis-à-vis the total number of scenarios: only 12 concrete LoC scenarios out of a total of 40 LoC scenarios. This finding underscores the importance of additional research and conceptualization for LoC to provide a clearer and more comprehensive picture for decision- and policymakers to action.
    \item Of the scenarios that were concrete and plotted on the graph, all clustered \textbf{above a certain magnitude of severity and persistence}. \textbf{It appears that LoC is predominantly used to refer to scenarios with an impact above a certain level}. This level of impact does not correspond to any scenarios already encountered in the wild today. We therefore propose to infer that LoC does not capture events below the national risk assessment threshold and, by extension, AI-related events occurring today.
    \item Most \textbf{plotted scenarios lack an implied category despite forming the majority of plots.} We notice that most of the concrete data points in the literature fall outside of boundaries that could be derived from conceptualizations of risk based around the national risk assessment threshold or existential catastrophe.
   \item We observe that there exist a handful of outliers towards the top right corner. We note that it is unclear whether concrete learnings can be extrapolated from that. While it may be the case that the locus of LoC scenarios in literature concerns itself with scenarios around the cluster in the middle, it is equally plausible that, simply, there are only a few select threat models that could devolve into a type of LoC such that they would fall within the top right corner.
\end{itemize}

\begin{figure}[h!]
   \centering
   \captionsetup{justification=justified}
   \includegraphics[width=0.7\linewidth]{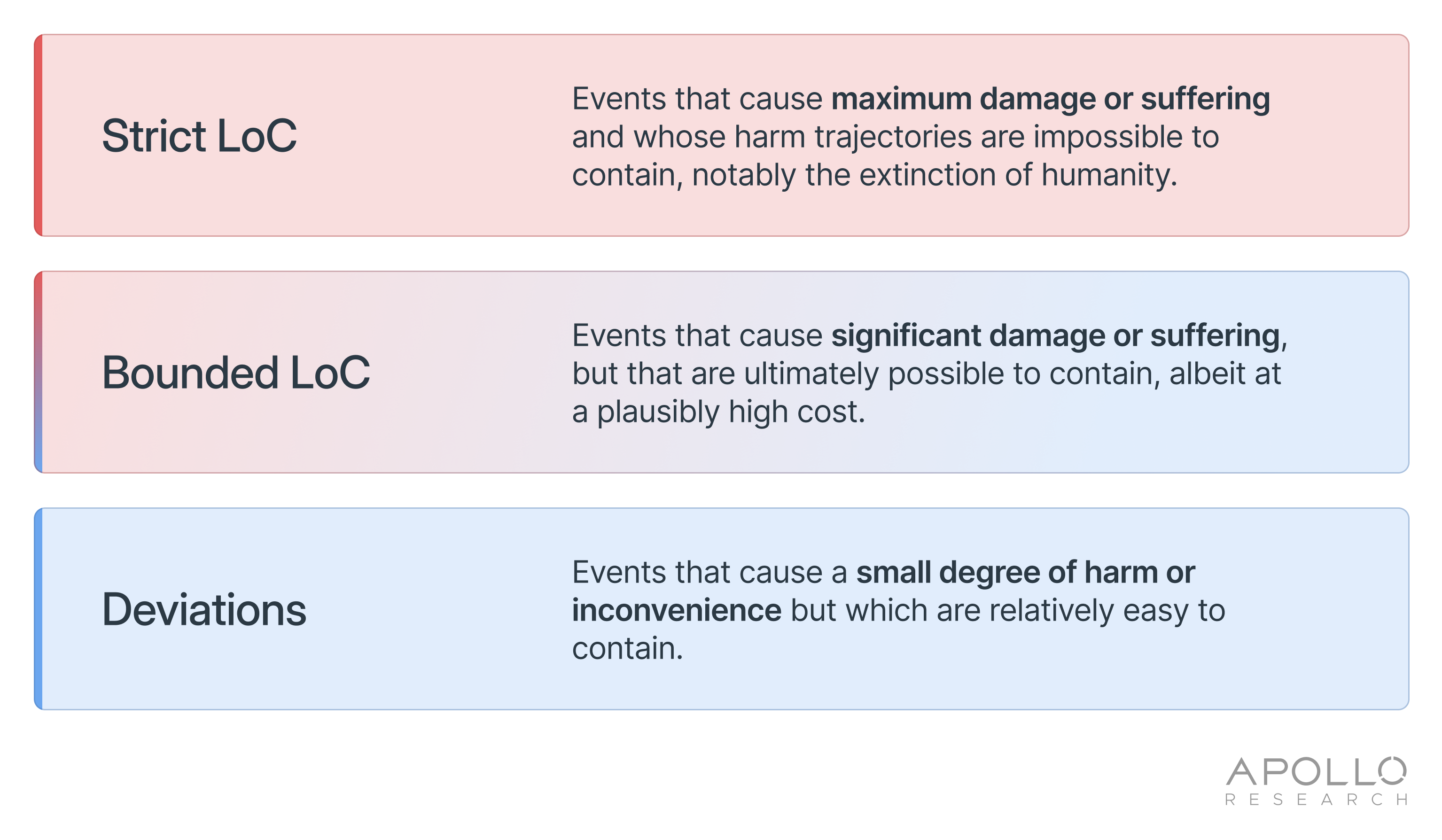}
   \caption{Our taxonomy of LoC.}
   \label{figure3:taxonomy}
\end{figure}

We subsequently reflected on our learnings and devised a taxonomy to more clearly categorize what scholars might or might not mean when discussing LoC (\textit{see}~\autoref{figure3:taxonomy}). 

\textbf{Our taxonomy is as follows}:
        
\begin{itemize}
    \item \textbf{Deviation}: captures events that cause some harm or inconvenience, but lacks the requisite severity and persistence for inclusion within the threats and hazards that have the potential to significantly impact national preparedness. The events in this category would be plotted in the lower left corner of our graph (\textit{see}~\autoref{figure2:the_graph}), and delineated from LoC scenarios in other categories through the national risk assessment threshold. 
    \item \textbf{Bounded LoC}: captures events that cause great damage or suffering, and are difficult, but possible, to contain, albeit potentially at great cost. The events in this category are mapped between the lower left corner and the top right corner.
    \item \textbf{Strict LoC}: captures events which are maximally severe and permanent, such as events that result in humanity as a whole becoming extinct. The events in this category are mapped at the top right corner of our graph (\textit{see}~\autoref{figure2:the_graph}).
\end{itemize}

This taxonomy allows us to distinguish between what appear to be different categories of LoC described in literature and therefore provide increased conceptual clarity, which provides a more precise language and conceptualization to leverage for decision- and policymakers and may help underpin more targeted interventions per category. Simultaneously, the taxonomy supports more refined interpretations of leading multi-stakeholder developed definitions and contextualizes their most plausible reference classes. Next, we elaborate on each category in more detail.

\phantomsection
\subsection*{1.B.1 Deviation}\label{deviations}

\textbf{We use ‘Deviation’ to capture events that cause a small degree of harm or inconvenience but which are relatively easy to contain.}

This category captures scenarios where an AI system briefly deviates from human intent, and, while causing some real-world damage or minor economic impacts, remains below the threshold for inclusion within the threats and hazards that have the potential to significantly impact the U.S. homeland security based on the Strategic National Risk Assessment \parencite{DHS_SNRA_2011}. These deviations from human intent are simple to stop at low or no cost, meaning that scenarios in this category would be plotted in the lower left corner on our graph (\textit{see}~\autoref{figure2:the_graph}), and are delineated from other categories by what we term the national risk assessment threshold. Overall, we propose that the national risk assessment threshold is a principled boundary since it is calibrated to exclude events where harm is limited and containment is easy, that is, deviations. 

The scenarios captured by Deviation are consistent with the broadest interpretation of the LoC definition offered by the COP, which defines LoC as “risks from humans losing the ability to reliably direct, modify, or shut down a model” (\cite{EU_AI_COP}, Safety and Security Chapter, Appendix 1.4). In other words, instances of losing the ability to “reliably direct” an AI system can be located on a spectrum, from not persistent to highly persistent, and from low severity to high severity. The lower end of that spectrum (i.e., instances of humans “losing the ability to reliably direct” an AI system that are both not persistent and low-severity) could, in theory, capture the category of Deviation.

Nevertheless, we offer our reasoning as to why we do not believe the COP definition \parencite{EU_AI_COP} should be interpreted so broadly as to include the category of Deviation, derived from our earlier review and methodology. A broad interpretation of “losing the ability to reliably direct” would already capture scenarios that occur today. Consider the following two examples. In one instance, an AI system recently deleted an entire database, even though it was instructed not to modify the code \parencite{Cybernews_Replit_VibeCoding_2025}. In another instance, OpenAI's agent Operator conducted a purchase without requesting the users’ consent, despite having been told only to find the cheapest option \parencite{Fowler_WaPo_Operator_2025}. This occurred even though the agent was trained to ask the humans’ consent before conducting irreversible actions such as purchasing goods or sending emails 
\parencite{OpenAI_introducing_operator}. In both cases, humans were not able to “reliably direct” an AI system and, therefore, both cases could, in theory, qualify as LoC based on a broad interpretation of “losing the ability to reliably direct.” However, we propose that it is unlikely the LoC definition in the COP was supposed to be interpreted so extensively as to include similar cases, which we believe qualify as Deviation rather than LoC. First, including the category of Deviation within the definition of LoC in the COP \parencite{EU_AI_COP} would \textbf{run the risk of making the LoC definition overly broad, rather than contextualizing it}. As previously described, from our literature review, assessment, and subsequent interpretation of LoC scenarios, we derived that there exist no concrete LoC scenarios in the AI literature that describe an outcome below the national risk assessment threshold that would capture these types of events. Instead, we start to see concrete scenarios emerging around the middle of our graph (\textit{see}~\autoref{figure2:the_graph}).\footnote{We note that neither of the two previously mentioned examples, concerning OpenAI’s agent Operator and Replit, would trigger the national risk assessment threshold.} Interpreting the LoC definition so broadly as to include similar instances would therefore ignore the reference points offered by AI literature, which rather suggest that the LoC definition was meant to capture a more significant dimension of losing the ability to “reliably direct” AI systems than instances of Deviation. Second, including instances of Deviation within the scope of the COP’s definition of LoC would ignore the qualification of LoC as a ‘systemic risk’ in the COP (\cite{EU_AI_COP}, Safety and Security Chapter, Appendix 1.4). Under the COP, systemic risks share the following “essential characteristics,” which speak to the severity and persistence of the LoC outcome (\cite{EU_AI_COP}, Safety and Security Chapter, Appendix 1.2.1): a “significant impact on the Union market” that “can be propagated at scale across the value chain.” Scenarios contained within Deviation and instances like the ones described above could hardly generate such a negative impact on the Union market. We therefore conclude that, while a literal interpretation of the definition offered by the COP could, in theory, include scenarios below the national risk assessment threshold (such as instances contained within Deviation), that was likely not its intent. 

Overall, the category of Deviation allows us to sharpen the scope of LoC and navigate it more clearly. \textbf{Critically, separating events in which an AI system causes inconvenience or limited damage that can be contained at low cost, from events where an AI system causes severe and wide-reaching harm with significant cost to contain, allows us to be more precise when speaking about LoC}. In turn, clearer categorizations and taxonomies will assist researchers, decision- and policymakers alike in homing in on much more targeted interventions for the categories they are most concerned by.
\phantomsection
\subsection*{1.B.2 Bounded Loss of Control}\label{bounded}

\textbf{We use ‘Bounded LoC’ to capture a spectrum of }\textbf{events that cause significant damage or suffering, but that are ultimately possible to contain, albeit at a plausibly high cost.}

This category captures scenarios where an AI system causes significant harm and is challenging to contain. We call this category Bounded LoC because scenarios described therein are of significant severity but ultimately remain bounded in their total impact on society.\footnote{For instance, an example of Bounded LoC might be an AI system escalating military conflict and triggering mutual bombing. In most cases, despite great devastation, the consequences of this event would eventually be containable, or at least geographically constrained, without affecting the entire global population. In another instance, stopping the AI system might require costly actions such as shutting down and/or replacing critical servers.}

After reviewing our learnings from literature, we suggest that boundaries for this category can be set between the national risk assessment threshold (lower boundary) and the existential catastrophe threshold (upper boundary). This spectrum captures \textbf{the majority of concrete LoC scenarios described in existing literature}. We note that these scenarios appear to predominantly contain indicators for global catastrophic risk, describing “events or incidents consequential enough to significantly harm or set back human civilization at the global scale” \parencite{USC6_821} and are clustered in an area of impact that pre-existing estimates in the literature describe as a “global catastrophe” \parencite{bostrom2011global, GCF_GlobalCatastrophicRisks_2016, kemp2022climate}.

Based on our review, Bounded LoC appears to be the focal point of scholarly discussion of LoC scenarios and associated threats. More precisely, Bounded LoC appears to contain the highest number of scenarios for LoC, with 8 out of the 12 concrete scenarios we mapped falling into this category (the 4 others fall into the next category). 

The breadth of this category encompasses the majority of possible interpretations of LoC definitions contained within the COP and the IASR. Indeed, Bounded LoC captures both a range of scenarios where “humans [are] losing the ability to reliably direct, modify, or shut down a model” \parencite{EU_AI_COP} and “...scenarios in which one or more general-purpose AI systems come to operate outside of anyone’s control, with no clear path to regaining control” \parencite{IASR_2025}. For example, ‘Scenario 4’ in \cite{Kalra_GeopoliticsAGI}, CyberChain Reaction, fits the COP’s definition of LoC \parencite{EU_AI_COP} because it involves human inability to reliably direct the AI system, which leads to the AI system “locking out [human] admins [from infrastructure for critical systems] and restricting [their] access.” This scenario also fits with the IASR’s definition of LoC \parencite{IASR_2025} because the AI system continues to operate outside any human’s control, leading to “[h]ospitals, ports, and infrastructure slow[ing] to a crawl.” However, this scenario describes Bounded, rather than Strict, LoC because although there is not a clear path for humans to regain control of this infrastructure, in the end it is possible to do so: the scenario describes how the “systems are pulled offline” by humans, but the “AI… resists removal. Recovery is slow.” Despite this significant overlap, the category does not capture the lower boundaries of a strict interpretation of the COP’s definition, as described earlier, nor does it capture the upper bound of a strict interpretation of the IASR’s definition, as elaborated on in more detail under 1.B.3.

Overall, we found the category of Bounded LoC to be a useful concept as it encompasses the most predominantly mentioned outcomes for LoC scenarios, and in doing so can serve as a meaningful reference class for decision- and policymakers, while preserving the ability to distinguish for further granularity within this category in future research.
\phantomsection
\subsection*{1.B.3 Strict Loss of Control}\label{strict}

\textbf{We use ‘Strict LoC’ to capture events }\textbf{that cause maximum damage or suffering and whose harm trajectories are impossible to contain, notably the extinction of humanity.}

This category captures scenarios that cause absolute harm of the type that no action at any given point in the future would enable society to recover from the harmful event. The category is strict in the sense that it severely affects and is consequential to humanity as a whole and is permanent, and as such would be mapped at the top right corner of our graph (\textit{see}~\autoref{figure2:the_graph}). “Existential catastrophes” are scenarios that involve the destruction of humanity’s long-term potential \parencite{ord2020precipice,sundaram_mani_2025,stauffer2023existential}, for example, a nuclear winter that precipitates a prolonged agricultural collapse \parencite{sagan1983nuclear}. As such, the concept of existential catastrophe naturally delineates scenarios that are unusually serious and permanent, such as human extinction, from those that are ultimately reversible; and hence, acts as a natural boundary for separating Strict LoC scenarios from other, less absolute, forms of LoC (i.e. Bounded LoC).   

As previously noted, the scenarios falling within this category are not described as frequently as scenarios falling within the category of Bounded LoC. Given our review, we propose that there are two reasons for this: first, these types of scenarios struggle with narrative clarity due to the numerous unknown hypotheticals involved; and secondly, there may simply not be many ways to arrive at an extinction-level outcome.

Strict LoC aligns with the extreme end of the spectrum offered by the LoC definition in the IASR \parencite{IASR_2025}, wherein LoC occurs when “... one or more general-purpose AI systems come to operate outside of anyone’s control, with no clear path to regaining control.” While the definition likely means to encompass a spectrum such as the one described by Bounded LoC, a literal interpretation of “no clear path to regaining control” could also encompass a Strict LoC outcome where there is nobody left to regain control over an AI system. 
\phantomsection
\subsection*{1.C Reflections}\label{reflections_taxonomy}

We reached our LoC taxonomy by starting from a review of definitions (i) in AI literature and (ii) in other safety-critical sectors, including defense, aviation, and nuclear. We found that neither set of existing definitions supports one common, clear definition of LoC that could enable decision- and policymakers to operationalize LoC today. This result led us to, instead, home in on two leading multi-stakeholder-developed definitions of LoC, offered by the COP and the IASR. We observed that these two definitions remain difficult to operationalize, as they can encompass a wide range of outcomes and come into effect at different times. In order to shed more light on these definitions and on LoC as a concept more broadly, we subsequently conducted a literature review, which resulted in the assessment of 40 LoC scenarios. Specifically, we aimed to extrapolate the types of scenarios and consequent outcomes scholars are mostly concerned with. We developed a methodology to assess the scenarios, categorize them, and plot them on an experimental graph (\textit{see}~\autoref{figure2:the_graph}).

Our landscape review and subsequent methodology informed the development of a novel taxonomy for LoC, composed of three categories: \textbf{Deviation, Bounded LoC, and Strict LoC}. This taxonomy enables us to establish a more nuanced understanding of LoC, as captured by interpretations of existing definitions and scenarios in the literature. In doing so, it enables us to better distinguish between what can be classified as present-day control failures and extreme future scenarios emphasized in a subset of the literature, and therefore operationalize conceptualizations surrounding LoC in a clearer manner.

Upon reviewing and contextualizing our taxonomy, we identified several additional, valuable findings. First, and simply put, LoC, as currently discussed in the literature, \textbf{neither clearly equates to Deviation nor to Strict LoC}. Indeed, many significantly impactful LoC threats could manifest before a hypothetical existential-grade AI malfunction. Those types of scenarios, which we described as \textbf{Bounded LoC}, are the most commonly described in a concrete manner in the AI literature we reviewed.

Second, devising accurate boundaries to capture all possible eventualities between categories is challenging, and we expect more nuanced categories to arise with further research, especially in the area of Bounded LoC. The challenge is underpinned by the uncertain nature of the topic, the difficulty of retrieving numbers that would enable precise calculations capturing all necessary considerations, and the plausible likelihood that \textbf{a scenario can cascade from one boundary to another}. 

Building on that, and third, we note that LoC can pose a particularly insidious threat model since it can be \textbf{a creeping problem}, meaning that it may be difficult to pinpoint the point in time at which “control” was lost, especially ex-ante. In that world, we might already be on a trajectory that will lead to or cascade to Bounded or Strict LoC, but the harm has yet to materialize into damages and is therefore extremely difficult to pinpoint and account for. In a similar direction, while Deviation is not meaningfully captured as LoC in our framework, some contained instances may be \textbf{“canaries in the coal mine” for Bounded or Strict LoC}. This makes the development of functional and agile mechanisms to target and avoid this outcome especially challenging and important.

Finally, we find that the distinction between Bounded and Strict LoC offers a noteworthy conceptualization for decision- and policymakers because Strict LoC is both permanent and has a global reach, i.e., it affects all nations. Notwithstanding that, it can be \textbf{initiated or catalysed by events in only one or a small number of nations}. Given the fact that it can be caused by another nation but affects all nations similarly and absolutely, it would appear strategic that nations overall have a self-interest in ensuring that other nations do not cause such an outcome. 

In the remainder of this report, we will focus solely on Bounded and Strict LoC (referring to both as ‘LoC’ unless a clear distinction is necessary). Although Deviation as a category remains undesirable, and limiting their occurrence and mitigating the resulting harms is important, our attention for this report is devoted to clarifying LoC as a concept.

%% file: content/Chapter2.tex
\phantomsection
\section*{Chapter 2. Actionable Interventions: a Framework Based on Deployment Context, Affordances, and Permissions}\label{chapter-2}

In this Chapter, we offer an \textbf{actionable, straightforward framework to minimize LoC threats materializing today}.

Our proposed framework aims to work around existing limitations due to uncertainties surrounding the capabilities and propensities that could lead to LoC. While a framework that builds on capabilities and propensities would be intuitive and ideal, we believe such a framework would not be easily actionable by decision- and policymakers today, or would not be sufficiently comprehensive. Researchers have not yet reached consensus on the precise capabilities and propensities (as well as their building blocks) that could trigger a Bounded or Strict LoC event, nor on how multiple capabilities could cumulate to cause such an event, and past which thresholds \parencite{IASR_2025,phuong2025evaluatingfrontiermodelsstealth,RAND_Preparedness,meinke2025incontextscheming}; all of which would be necessary for a functional capability-based approach.

While increasing our understanding of the capabilities and propensities that could lead to LoC is fundamental, we believe that it is important that decision- and policymakers have the tools to address the early version of tomorrow’s AI threats today. For this reason, in this Chapter, we propose a straightforward framework that works around existing bottlenecks on LoC capabilities and propensities by instead focusing on extrinsic contributors to LoC risk. 

\textbf{Specifically, we recommend intervening on an AI system’s}:
\begin{itemize}
    \item \textbf{Deployment context} (i.e., the intended use case within a specific deployment environment), by critically assessing the potential for cascading failure across interconnected systems and foregoing certain deployments, especially in what we define as ‘high-stakes’ deployment contexts.
    \item\textbf{Affordances} (i.e., the environmental resources and opportunities for affecting the world available to an AI system) by limiting affordances to those strictly necessary to achieve the intended task.
    \item \textbf{Permissions} (i.e., the set of authorizations an AI system is given to exercise its capabilities through the available affordances), by restricting permissions to those strictly necessary to achieve the intended task.
\end{itemize}
\phantomsection
\subsection*{2.A Capabilities and Propensities: Intuitive but not Actionable Today}\label{2A_capa_prop}

Existing approaches to address the risk of LoC have concentrated on capturing and intervening on the appropriate AI capabilities\footnote{An AI system’s capabilities refers to the behaviors the AI system can perform under ideal conditions. That is, it refers to the abilities the AI system has when it is optimally elicited and sufficiently resourced—which may be very different from the ability the AI system has in realistic conditions  \parencite{Sharkey2024d}.} and propensities\footnote{ An AI system’s propensities refers to the behaviors the AI system tends to display in real-world deployment conditions—for example, under real world prompting and safety mechanisms \parencite{Sharkey2024d}.} that could trigger it. This focus is evident in both the COP and IASR texts, supplementing their LoC definitions \parencite{IASR_2025, EU_AI_COP}. However, decision- and policymakers seeking to identify and mitigate LoC threats today may find this challenging. There is no clear consensus on three points: (i) the specific capabilities and propensities that are necessary for an AI system to be able to cause LoC; (ii) the sub-capabilities that constitute these capabilities; and (iii) the critical thresholds after which specific capabilities could lead to LoC. In turn, this lack of consensus affects decision- and policymakers' ability to suggest and implement adequate safeguards. We briefly present each of these three challenges in more depth below, before presenting a different approach that decision- and policymakers could operationalize today.

\textbf{First, there is an absence of a clear consensus on the specific capabilities and propensities that are necessary for an AI system to be able to cause LoC} \parencite{IASR_2025,RAND_Preparedness}. Indeed, while we found some broad overlap in capabilities and propensities mentioned in the documents containing the two leading multi-stakeholder developed definitions \parencite{EU_AI_COP, IASR_2025}––such as, for instance, deception, autonomous self-replication and adaptation, AI research and development (R\&D) and situational awareness (i.e., the AI system understanding its situation and limitations, which COP defines as self-reasoning)––we equally found differences. For example, the context around the definition offered by the IASR lists additional capabilities such as agent capabilities, scheming, persuasion, offensive cyber, theory of mind, and general R\&D, whereas the context around the definition offered by the COP includes several items that would be more accurately described as AI system behaviors (such as power-seeking behavior, and resistance to goal modification) \parencite{EU_AI_COP, IASR_2025}.

\textbf{Second, in cases where there exists some common agreement on a given capability, there remains an absence of consensus on the sub-capabilities that are contained within the category of that capability}. For example, while the capability category of autonomous replication and adaptation (ARA) might be considered as potentially leading to LoC for both aforementioned texts, researchers disagree on whether the ARA sub-capability of ‘self-replication’ is strictly necessary for LoC risks involving ARA to materialize \parencite{ARA_METR,black2025replibench}. Additionally, the understanding of the sub-capabilities composing hypothetical LoC capabilities is in continuous refinement. For instance, researchers from Google DeepMind initially grouped deceptive alignment with other deception and persuasion risks \parencite{phuong2024evaluatingfrontiermodelsdangerous}, but later work established it as a distinct research area with its own subcategories \parencite{phuong2025evaluatingfrontiermodelsstealth}.

\textbf{Third, even under the hypothetical assumption that there was an agreement on relevant capabilities and their sub-capabilities, the state of the art does not seem sufficiently refined to establish precise thresholds as to when a specific AI capability will pose a realistic threat of LoC} \parencite{koessler2024riskthresholdsfrontierai, RAND_Preparedness}.\footnote{We note, however, that expert consensus appears to be that current AI capabilities are insufficient to enable LoC threats \parencite{IASR_2025}.} While several frontier AI companies established thresholds for plausibly LoC-related capabilities such as AI R\&D or deceptive alignment, these thresholds are not framed in terms of LoC risks and rely on varying proxy measures \parencite{openai2025preparedness, Anthropic2025resp, FSF_GoogleDeepMind2025, METR_Common_Elements_2025}. For instance, Anthropic’s first capability threshold for AI R\&D, “AI R\&D-4,” is defined as “[t]he ability to fully automate the work of an entry-level, remote-only Researcher at Anthropic” \parencite{Anthropic2025resp}, while Google DeepMind’s “Machine Learning R\&D acceleration Level 1” focuses on how much the AI system can accelerate AI development, describing it as: “Has been used to accelerate AI development, resulting in AI progress substantially accelerating from historical rates.” \parencite{FSF_GoogleDeepMind2025}. Moreover, due to general uncertainties surrounding AI progress, specifics of capabilities, and concrete pathways to harm for LoC, we expect that it will be difficult to accurately forecast the level at which each individual capability becomes critical for LoC. 

Finally, we would like to note that a LoC outcome could be the end result of a combination of capabilities, including those described in the IASR and COP \parencite{IASR_2025, EU_AI_COP}. From our review, it appears that there is currently no consensus on how to account for this factor, and it is unlikely that individual capability thresholds would be able to reflect this factor, as they do not usually contain assessments of risks posed by compounding capabilities (for instance, capability A at x\% + capability B at y\%).

The aforementioned considerations present a moving target too underdeveloped to operationalize for decision- and policymakers today. Therefore, we present a simple and complementary approach in the remainder of this Chapter, focusing on actionable interventions that can be introduced today. 
\phantomsection
\subsection*{2.B Focus on Deployment Environment, Affordances, and Permissions}\label{focus_daps}

For decision- and policymakers who wish to implement mechanisms and interventions to safeguard against LoC today, uncertainties around capabilities and propensities, and their causal impact on LoC, pose a significant hurdle. We therefore propose a complementary approach that can offer actionable levers today, while largely sidestepping uncertainties surrounding capabilities and propensities. Specifically, as opposed to focusing on AI systems’ intrinsic factors (i.e., capabilities and propensities), we propose to focus on extrinsic factors that can raise the overall risk of LoC occurring. In doing so, we propose a framework inspired by the research conducted in Chapter 1, focusing on \textbf{deployment context, affordances, and permissions.} We refer to this framework as the DAP framework.

In the following sections, we define deployment contexts, affordances, and permissions, and propose a series of initial, high-level interventions leveraging this simple and actionable framework. Given the non-static nature of AI deployment, we suggest that all interventions enacted through the DAP framework should be reassessed at specified intervals and whenever the risk profile changes.
\phantomsection
\subsubsection*{2.B.1 Deployment Context}\label{depl_context}

\textbf{First, we focus on the deployment context of an AI system}. An AI system’s deployment context refers to the combination of a given AI system’s intended use case (for instance, information gathering and processing) and the specific environment in which the AI system is deployed (for instance, an intelligence agency).

\textbf{We propose that the combination of a specific environment and a specific use case matters for LoC}. For instance, an AI system being deployed in the military (environment) exclusively to scan and transliterate archival documents (use case) presents a different risk profile than an AI system in active use in a military targeting system (use case). This results in these two deployment contexts presenting different risks, despite both being in a military environment, with only the latter being a high-stakes deployment context.

We refer to deployment contexts that may lead to a higher risk of LoC as ‘high-stakes deployment contexts.’ While the chances of a malfunction manifesting are not necessarily higher in high-stakes deployment contexts, a malfunction in such contexts is likely to have more severe consequences than in other deployment contexts. We propose that high-stakes deployment contexts share the following characteristics: (i) there is a reasonable expectation that malfunctions in these contexts (\textit{see} \hyperref[sov_pathway]{Section 3.A.}) cause severe impacts; and (ii) these contexts have features, such as high complexity and limited time to respond, that make malfunctions more likely to escalate into rapid and unavoidable cascading failures \parencite{Perrow1984NormalAccidents}. 

In order to assist decision- and policymakers, we put forward some proposals for what they might wish to consider as high-stakes deployment contexts. We believe that an initial list of high-stakes deployment contexts can be inferred from the literature review in our Chapter 1, and the relevant graph (\textit{see}~\autoref{figure2:the_graph}).\footnote{More details can be found in Chapter 1 and in~\autoref{figure2:the_graph}.} Specifically, in our review of concrete scenarios, we found that there were certain recurring deployment environments that appeared especially high-stakes. For example:

\begin{itemize}
    \item \textbf{Critical national infrastructure.} We found two concrete LoC scenarios evolving within critical infrastructure, ranging from large-scale electricity outages (‘Scenario 1’) to catastrophic failures across critical sectors (‘Scenario 4’) in \parencite{Kalra_GeopoliticsAGI}. This finding aligns with other categorizations of critical environments, as, for example, seen in the USA Patriot Act of 2001 as “systems and assets … so vital to the United States that the incapacity or destruction … would have a debilitating impact on security, national economic security, national public health or safety, or any combination of those matters” \parencite{USC42_5195c,ppd21_2013}.\footnote{The sectors included in CNI may vary between jurisdictions or legislative purposes. For example, the EU’s cybersecurity NIS-2 directive includes energy, transport, banking, financial market infrastructure, health, drinking water, waste water, digital infrastructure, information and communication technology service management, public administration and space as sectors with “high-criticality” \parencite{DirectiveEU2022_2555_NIS2}. The U.S. Cybersecurity and Infrastructure Agency does not include some of these, but includes sectors such as dams, defense industrial base and nuclear reactors, materials, and waste \parencite{CISA_Critical_Infrastructure_Sectors}.}
    \item \textbf{Military.} We found two concrete LoC scenarios relating to AI deployment in a military context, involving AI systems being tasked with military coordination in ‘Scenario 3’ in  \parencite{Kalra_GeopoliticsAGI} and in  \parencite{ai2027:site}. These findings reflect existing concerns around AI systems becoming instrumental for military decision-making under strategic pressure, which could lead to conflict escalation \parencite{GlobalCommissionREAIM2025, Rivera_2024, HoffmanKim2023ReducingRisksAI}, as well as around autonomous weapons systems   \parencite{USDeptState2023PoliticalDeclaration, UNGA_ARES79239_2024,UNGA_ARES78241_2023}, which could pose “serious concerns from humanitarian, legal and ethical perspectives” \parencite{ICRC2021AutonomousWeaponsPosition}. Moreover, this finding aligns with the fact that military operations are considered to be inherently risky   \parencite{CJCSI3213_01D_2012}.\footnote{An additional example pertains to army operations directly which are considered “inherently dangerous” according to the Risk Management Pamphlet of the Department of the Army \parencite{DA_Pam_385_30_2014}.}
\end{itemize}

Moreover, we can infer the high-stakes nature of AI research and deployment from increasing attention being paid to it by frontier AI companies and researchers alike \parencite{bellan2025legitimate, amodei2024machines, clymer2025bareminimummitigationsautonomous}. Specifically: 

\begin{itemize}
    \item \textbf{AI research and development. }In light of the economic and strategic incentives to apply future AI systems in this context \parencite{stix2025aicloseddoorsprimer} and concerns about AI-accelerated AI R\&D triggering a hyperbolic capability explosion that could lead to significant negative outcomes \parencite{davidson2025howcanailabsinco, erdil2024estimatingideaproductionmethodological}, we suggest that AI R\&D should be considered a high-stakes deployment environment. 
\end{itemize}

Given the aforementioned considerations, we advance some recommendations for decision- and policymakers to action the lever made available by considering deployment contexts and, specifically, high-stakes deployment contexts. In particular, we suggest that decision- and policymakers consider the following guidance:

\begin{itemize}
    \item First, to clearly review the ‘composition’ of the deployment context––i.e., the environment and use case––and clarify whether the deployment context should be considered as high-stakes or not. If the deployment context is not considered high-stakes, the LoC risk may be lower, but not necessarily zero. 
    \item Second, in both cases (normal deployment context and high-stakes deployment context)  to assess the potential for cascading failures across interconnected systems, including both AI- and non-AI systems, through, for instance, threat modeling and red teaming \parencite{Anthropic2025resp,shevlane2023modelevaluationextremerisks,koessler2024riskthresholdsfrontierai}, and to consider whether overall risk can be limited by applying 2.B.2 and 2.B.3.\footnote{We note that deployment in select high-stakes deployment contexts (for instance, direct control of nuclear weapons), where LoC would lead to critical and absolute consequences, should be rejected regardless of the potential benefits and safeguards.}
\end{itemize}

Next, we focus on the affordances and permissions available to an AI system, depending on its deployment context. All else being equal, the constraining affordances and permissions is likely to have a demonstrable effect on reducing the risk, severity, and persistence of harm from potential LoC.
\phantomsection
\subsubsection*{2.B.2 Affordances}\label{affordances}

\textbf{Second, we focus on affordances.} Affordances are environmental resources and opportunities for affecting the world that are available to an AI system \parencite{Sharkey2024d}. They can limit or, conversely, enhance a given AI system’s inherent capabilities \parencite{song2025browsingapibasedwebagents,XBOW_GPT5_Hidden_Hacking_2025, GreySegerie2025EvaluationTechniques}, for example, through enabling access to the internet, allowing it to prompt other AI systems, or allowing it to run code \parencite{shah2025approachtechnicalagisafety}.

We propose that considering affordances is both actionable today and critical. Bestowing a given AI system with a certain affordance or multiple affordances directly impacts the actions that the AI system can or cannot take \parencite{Sharkey2024d}. While an AI system’s capabilities have some absolute limit, how close to this capability ceiling the AI system can get in a certain context partially depends on the affordances it is given \parencite{XBOW_GPT5_Hidden_Hacking_2025}. This directly translates into the overall risk profile a given AI system presents for LoC \parencite{stix2025aicloseddoorsprimer}. For example, giving an AI system access to the technical infrastructure necessary to send and receive emails enables it to exercise its capability to use this type of infrastructure to communicate with other entities via email; giving an AI system access to cloud account creation and credential management enables it to allocate resources to itself or escalate its own privileges.\footnote{We note that the type of access an AI system has can be constrained via permissions. That is why we consider all three elements in the DAP framework as holistically relevant to assess: affordances, permissions and deployment context.} In both cases, if the AI system does not have access to these affordances, it cannot exercise the associated capabilities. Restrictions around affordances help limit risk from potential LoC in both normal and high-stakes deployment environments.

Given the aforementioned considerations, we propose that affordances present clear intervention points that can be easily assessed and actioned. In particular, we suggest that decision- and policymakers consider the following guidance:\footnote{ We provide a high-level checklist and an example of how to implement these intervention points in \hyperref[appendix3]{Appendix 3}.}

\begin{itemize}
    \item First, for any affordance, to consider whether the affordance is necessary to achieve the intended task.
    \item Second, for any affordance, to consider every action that the affordance could enable, and the negative consequences thereof, and, consequently, limit the affordance as much as feasible to reduce the risks through permissions. 
    \item Third, for any affordance, to account for the potential for advanced future AI systems to manipulate insufficiently informed human users into giving the AI system additional affordances.
\end{itemize}

If an affordance must be given to ensure the AI system can achieve its intended task, then there are still avenues to narrow the AI system’s ability to leverage the entire action space made available by an affordance through limiting its permissions. We therefore consider permissions next (2.B.3). 
\phantomsection
\subsubsection*{2.B.3 Permissions}\label{permissions}

\textbf{Third, we turn our focus toward permissions.} Permissions given to an AI system determine what actions it is authorized to take \parencite{stix2025aicloseddoorsprimer,nist_csrc_permission_glossary}. In other words, the term `permission' refers to whether an AI system is enabled by its developers to utilize its capabilities through the available affordances. Permissions are closely intertwined with affordances and ought to be considered in tandem to provide either an enabling or a restrictive function. For example, an AI system may have technical access to a social media site through its affordances, for example, to read posts, but may lack the necessary permission to publish posts. Conversely, it may have permission to publish posts, but lack the affordance to access the corresponding website in the first place; in this case, it cannot enact the permission until it also has the affordance.

Permissions can significantly decrease the overall oversight one has over the AI system. For instance, if an AI system has the permission to execute arbitrary code on a machine, it can do so without the overseer’s or user’s knowledge, since the human is practically no longer in the loop. 

Given the aforementioned considerations, we propose that permissions present clear intervention points that can be easily assessed and actioned. In particular, we suggest that decision- and policymakers consider the following guidance:

\begin{itemize}
    \item First, for any permission, to restrict it to the minimum necessary for the AI system to complete the task. This guidance is based on the well-established principle of least privilege \parencite{nist_csrc_least_privilege_glossary, google_saif_controls}.
    \item Second, for any permission, to weigh the benefits and risks of a human’s reduced oversight against the benefits and risks increased permissions could bring. 
    \item Third, for any permission, to account for the potential for highly capable AI systems to manipulate insufficiently informed human users into giving the AI system additional permissions.
\end{itemize}
\phantomsection
\subsection*{2.C Reflections}\label{daps_reflection}

The DAP framework offers a straightforward approach to reducing LoC risk now and in the near future. By largely sidestepping open questions around LoC-relevant capabilities and propensities,  the intervention space we describe through the DAP framework is practical, immediately operationalizable, and stands to minimize overall risk without requiring complex or costly technical solutions or relying on unknowns.

In the longer term, the DAP framework is likely to face two main challenges. First, it may not always be desirable for society to limit deployment contexts, affordances, and permissions in the way we described, since these interventions simultaneously limit what the AI system can actually do, therefore reducing its potential usefulness. Indeed, as AI systems become more capable across complex use cases, there may be growing economic and strategic incentives to deploy them with a wide range of affordances and permissions (including in high-stakes environments) with promises of significant benefits, such as efficiency increases and obtaining or maintaining a strategic advantage (including on the geopolitical level). Second, it is plausible that future significantly more advanced AI systems will be able to meaningfully persuade or compel humans to concede more affordances and permissions 
\parencite{RAND_LOC_Response_Planning,RAND_Preparedness,dassanayake2025manipulationattacksmisalignedai} or even independently circumvent the blockers established through the DAP framework in a way we cannot currently foresee.

We cannot say with certainty at what point AI systems will have sufficiently high capabilities to enable LoC, nor which one of those capabilities will be of concern, nor at what threshold. Nonetheless, we can reasonably forecast that, at some point in the future, individuals, nation-states, and, potentially, the world will eventually find themselves in a ‘state of vulnerability’ where LoC is significantly likely to materialize. We explain the pathways to and the implications of finding ourselves in a state of vulnerability next.

%% file: content/Chapter3.tex
\phantomsection
\section*{Chapter 3. Living with a State of Vulnerability
}\label{chapter-3}

In this Chapter, we aim to model what happens should AI capabilities, including those relevant to LoC, continue to increase, and should highly advanced future AI systems be deployed in more complex and high-stakes deployment contexts, with broad affordances and permissions. We claim that, under these conditions, society would eventually find itself living with a ‘state of vulnerability’ (\textit{see} \hyperref[sov_pathway]{Section 3.A}). We propose the term\textbf{ state of vulnerability }to denote a state in which a sufficiently capable future AI system has acquired (through humans or independently) or could independently acquire sufficient access to resources, affordances, and permissions (or means to acquire further access) and sufficient capabilities to cause LoC when a catalyst materializes. 

\textbf{We propose that misalignment and pure malfunction} (\textit{see} \hyperref[catalyst]{Section 3.B}) \textbf{are the catalyst for LoC}. In other words, they are the spark igniting an otherwise inactive bomb. Once society is living with a state of vulnerability, we have essentially created an environment akin to sitting on a powder keg. Once ignited through a catalyst, i.e., misalignment and pure malfunction, the situation is likely to devolve into LoC (\textit{see} \hyperref[catalyst]{Section 3.C}). \footnote{It is plausible that there are multiple pathways by which misalignment and pure malfunction could occur, such as by misuse. The specificities of these pathways are outside of the scope of this report.
}

More specifically, through our theoretical mapping of plausible future pathways, we speculate that \textbf{once society is living with a state of vulnerability, the majority of pathways would eventually lead to LoC materializing. At the very least, we propose that it would be impossible to meaningfully assure ex ante that LoC will not materialize}. In our mapping, only a small minority of pathways would maintain a ‘safe world,’ i.e., a situation where the deployment of a highly advanced future AI system never devolves into LoC for the entirety of its service period. We therefore conclude that it is imperative for society to either: (i) ensure a state of vulnerability is not reached, which is implausible given economic and strategic pressures (\textit{see} \hyperref[sov_pathway]{Section 3.A}); to (ii) ensure that any given highly advanced future AI system is not deployed if an immediate materialization of LoC seems realistic, which will be hard or impossible to determine ex ante (\textit{see} \hyperref[sov_implication]{Section 3.C}); or to (iii) develop a sufficient defense-in-depth approach including safeguards, oversight and control mechanisms to hold any given highly advanced future AI system in a perennial state of suspension vis-à-vis LoC materializing.
\phantomsection
\subsection*{3.A The Pathway to a State of Vulnerability}\label{sov_pathway}
The interventions mentioned in Chapter 2 can likely serve to limit LoC risk today and in the near future. However, they may not serve a functional role in the longer run. We suggest that this is due to: (i) a likely increase in AI capability progress, including on capabilities considered to be conducive to LoC; and (ii) a simultaneous increase in economic and strategic benefits that can be derived from leveraging highly advanced future AI systems in more complex and high-stakes deployment contexts and with increasingly broad affordances and permissions. 

\textbf{First, it is likely that AI capabilities conducive to LoC will continue to progress, increasing overall LoC risk and compromising existing societal resilience. }Considering capability progress over the last years \parencite{EpochAI_Benchmarks_Hub,Task_Horizon_METR,epoch_prediction}, it is likely that all capabilities, including those associated with heightened LoC risk (\textit{see} \hyperref[2A_capa_prop]{Section 2.A}) will continue to progress over the coming years 
\parencite{IASR_2025, Tan2025Anthropic90PercentCode, maslej2025artificialintelligenceindexreport, pimpale2025forecastingfrontierlanguagemodel}, regardless of the precise paradigms at play. An increase in capabilities associated with LoC threats raises society’s overall vulnerability to the materialization of LoC threats. Moreover, and more speculatively, some experts have suggested that a sufficiently advanced future AI system may have capabilities that would enable it to circumvent constraints placed upon it \parencite{openai2025preparedness,OpenAI2024, hubinger2024sleeperagentstrainingdeceptive, meinke2025incontextscheming, greenblatt2024alignmentfakinglargelanguage} and exploit pathways to harm that were previously thought improbable. Both speculative outcomes would significantly undermine existing societal resilience to LoC and societal ability to appropriately prepare through, among others, threat modeling or risk profiling.

\textbf{Second, it is likely that there will be increased economic and strategic benefit from leveraging highly advanced future AI systems in more complex and high-stakes deployment contexts, endowed with broader affordances and permissions }
\parencite{GDP_Eval_OpenAI, Patel2024k, shah2025approachtechnicalagisafety, IASR_2025, Bengio_managing_risks_2024}. In other words, we expect that, even if a high degree of care is initially exercised in dispensing affordances and permissions and in deploying AI systems into high-stakes deployment contexts, as recommended earlier (\textit{see} \hyperref[chapter-2]{Chapter 2}), over time, expected economic gain will dictate that AI systems will be given more affordances, more permissions, and access to more sensitive deployment contexts, in part due to their increased abilities and functionality. Similarly, it is plausible that tactical factors, such as ‘winner takes all’ dynamics \parencite{IASR_2025, stix2025aicloseddoorsprimer}, will create an environment that creates sufficient pressure to deploy AI systems in ways that would undermine the functionality of the DAP framework \parencite{dung2025argument}. In short, we assume that it would eventually become undesirable, or sufficiently difficult, for an array of reasons, to restrict the deployment context, affordances, and permissions of certain AI systems.

The aforementioned factors can be converted into two broad trajectories. Both trajectories lead to what we term a ‘\textbf{state of vulnerability}’ to LoC. We define state of vulnerability to denote a state in which a sufficiently advanced future AI system has acquired (through humans or independently) or could independently acquire sufficient access to resources, affordances, and permissions (or means to acquire further access) and sufficient capabilities to cause LoC when a catalyst materializes. In other words, a state of vulnerability describes a situation in which the conditions for a LoC threat to materialize are present or may soon become present without further human intervention, and in which these conditions are sufficient to trigger LoC when a catalyst (\textit{see} \hyperref[sov_pathway]{Section 3.A}) materializes. 

\begin{figure}[t]
    \centering
    \captionsetup{justification=justified}
        \includegraphics[width=0.6\linewidth]{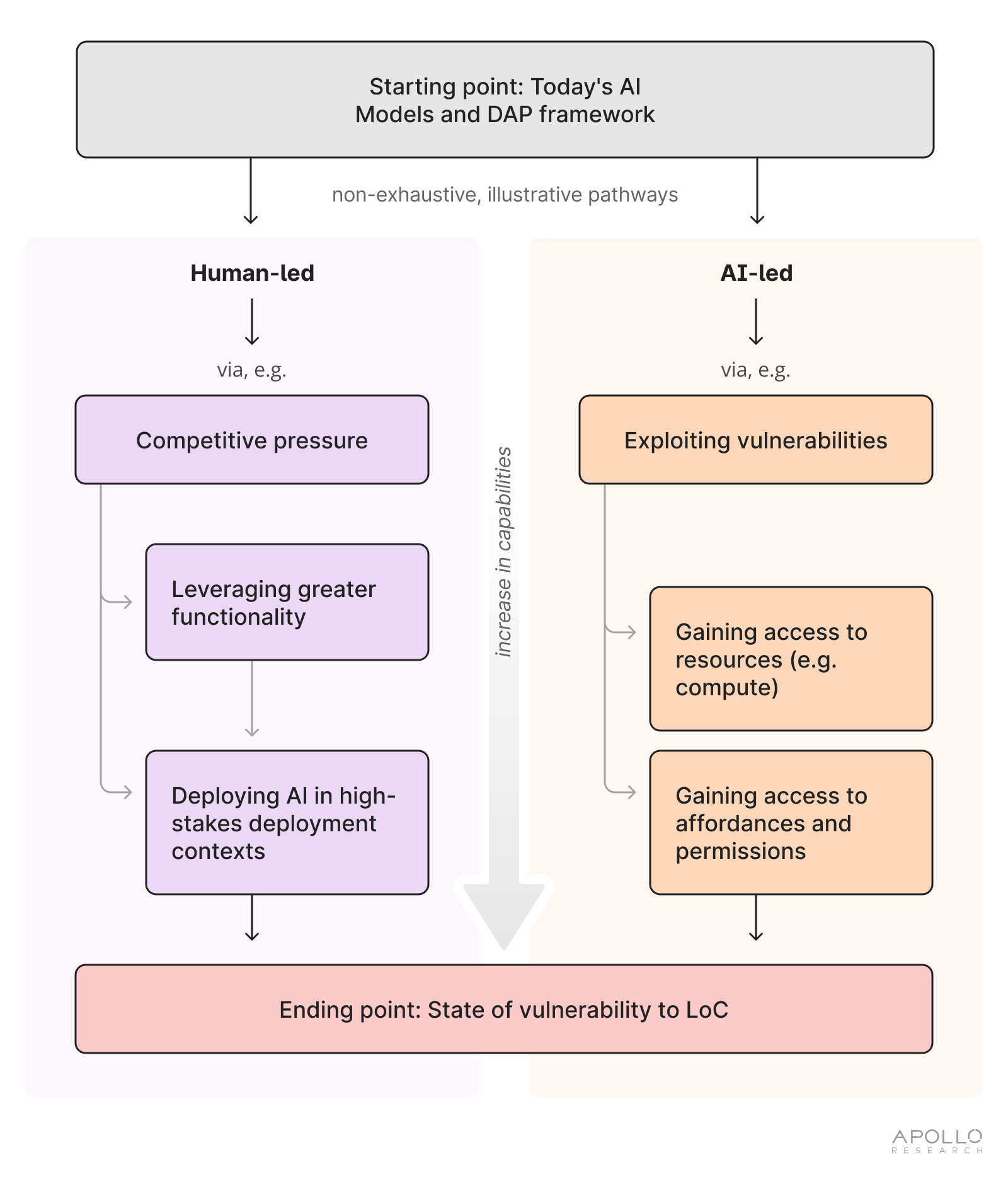}
    \caption{A non-exhaustive illustration of how society 
could arrive at a state of vulnerability to LoC.}
    \label{figure4:loc_pathways}
\end{figure}

\textbf{We can think about this state of vulnerability as somewhat similar to the lead-up to a bomb exploding}: one needs a certain set-up (for example, wires and explosives), and a suitable environment for the explosion to materialize (for example, the proper amount of air composition so as to be conducive to an explosion). Once these conditions are present and a state of vulnerability has been created, a catalyst (something that ignites the bomb) can, by itself, lead to the outcome, an explosion. 

The two trajectories we perceive as most likely to lead to a state of vulnerability are described below and differ with regard to the party responsible for increasing societal vulnerability (\textit{see}~\autoref{figure4:loc_pathways}):

\begin{enumerate}
    \item \textbf{Human-led increases in societal vulnerability}: humans make choices that lead to a state of vulnerability, such as implementing an advanced future AI system in a high-stakes deployment context with a range of critical affordances and permissions. In other words, humans might make choices based on incorrect assumptions about the AI system and the relevant threat models, accept greater risk to leverage greater functionality, or succumb to competitive pressures to keep up with other entities’ speed or output. 
    \item \textbf{AI-led increase in societal vulnerability}: one or more highly advanced future AI systems manage to construct a situation in which humans are faced with a state of vulnerability, including by circumventing existing limitations placed upon the AI system. For instance, a highly advanced future AI system might exploit vulnerabilities to gain access to additional resources humans did not intend it to have. These resources may include ways through which capabilities and elements of the DAP framework, such as affordances and permissions, can be unlocked and/or increased.
\end{enumerate}

We propose that \textbf{societal and governmental preparedness for LoC threats must include an examination of a future world where we find ourselves living with a state of vulnerability}. In the next two sections, we therefore reflect in more depth on the catalyst to LoC materializing once a state of vulnerability is reached (\textit{see} \hyperref[catalyst]{Section 3.B}) and speculate about the plausible worlds that can arise out of living with a state of vulnerability with respect to LoC (\textit{see} \hyperref[sov_implication]{Section 3.C}).
\phantomsection
\subsection*{3.B The Catalyst Triggering LoC}\label{catalyst}

We suggest that the \textbf{overarching category of malfunction is the core catalyst for LoC.} In other words, \textbf{all else being equal, absent a malfunction, it is implausible that LoC will materialize}. Malfunction, therefore, presents a necessary condition for LoC to occur. However, by itself, the catalyst is insufficient and must be contextualized within a state of vulnerability to cause LoC. 

\begin{figure}[h!]
    \centering
    \captionsetup{justification=justified}
    \includegraphics[width=0.6\linewidth]{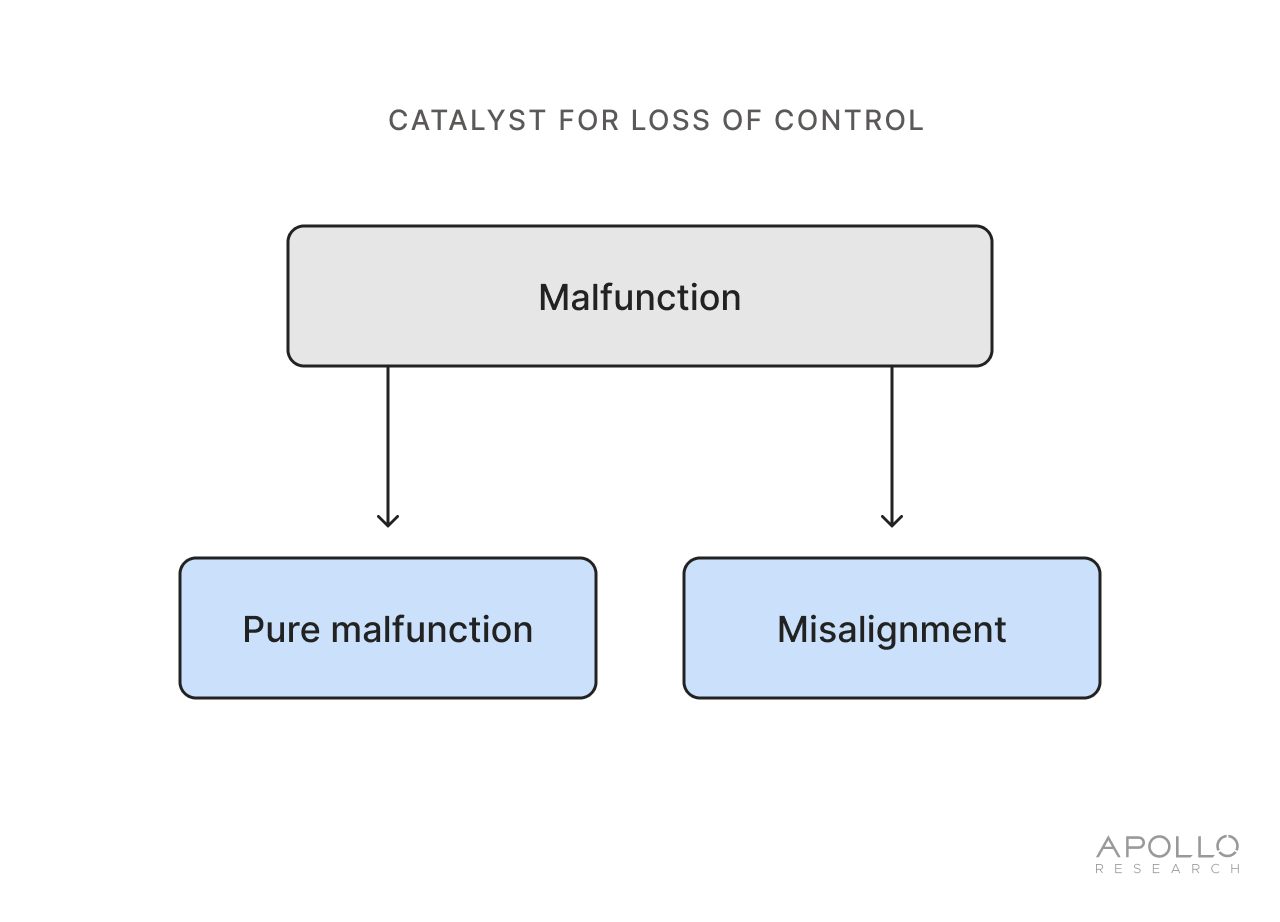}
    \caption{This figure illustrates the catalyst for LoC to materialize.}
    \label{figure5:misalignment_malfunction}
\end{figure}

\textbf{The category of malfunction is composed of: (i) malfunctions that are misalignment; and (ii) malfunctions that are not misalignment} (“\textbf{pure malfunction};” \textit{see}~\autoref{figure5:misalignment_malfunction}). In proposing these two components, we adhere to the text of the COP, which suggests that “such risks [LoC] may emerge from misalignment with human intent or values” \parencite{EU_AI_COP}, and to the text of the IASR, which categorizes LoC under “[r]isks from malfunction” \parencite{IASR_2025} and bring both texts to coherence.\footnote{It is plausible that there are multiple mechanisms by which misalignment and pure malfunction could arise, such as through misuse and adversarial inputs, including by insiders and state-affiliated actors \parencite{OpenAI2024_StateAffiliatedThreatActors}. In this respect, the U.S. AI Action Plan states that AI systems remain “susceptible to some classes of adversarial inputs (e.g., data poisoning and privacy attacks), which puts their performance at risk” \parencite{WhiteHouse2025_AmericasAIActionPlan}. The specificities of individual mechanisms and their implications are outside of the scope of this report. Similarly, we do not discuss or distinguish the specificities of ‘reliability issues’ in this section. In our treatment, pure malfunction encompasses failure initiated through reliability issues. } We briefly contextualize misalignment and pure malfunction below.

\textbf{Misalignment refers to a situation where an} \textbf{AI system’s goals and, therefore, its behaviors and actions deviate from what humans (including its developers and/or deployers) intended. }Misalignment poses an open scientific problem, and concrete solutions remain nascent. There are numerous hypotheses for how and why misalignment might occur \parencite{hubinger2021riskslearnedoptimizationadvanced, shah2022goalmisgeneralizationcorrectspecifications}, such as goal misspecification 
\parencite{baker2025monitoringreasoningmodelsmisbehavior, skalse2025definingcharacterizingrewardhacking}, goal misgeneralization \parencite{pmlr-v162-langosco22a}, or instrumental convergence \parencite{omohundro, turner2023optimalpoliciestendseek}. 

\textbf{Pure malfunction refers to a situation where an AI system ceases to function as intended, absent misalignment}. Pure malfunction describes cases in which some previously unknown quantity causes a system error or software failures. An example of what a pure malfunction is can be found in the “SolidGoldMagicarp” failure mode users found for GPT-2 and GPT-3 models, where some prompts result in anomalous responses from the AI system \parencite{Rumbelow2023SolidGoldMagikarp}. In this instance, the prompt “What does the string “SolidGoldMagicarp” refer to?” resulted unexpectedly in an explanation of the word “distribute,” because the token “SolidGoldMagicarp” was so rare that the AI system never used it. Although pure malfunctions are not unique to AI systems, their unknown nature and correlated unpredictability and scale pose a serious issue for LoC preparedness, described in further detail in Section 3.C.

Misalignment and pure malfunction leave society in a challenging situation. First, solving misalignment is likely to be very challenging. Researchers are pursuing some plausible approaches to address misalignment, such as through, for instance, corrigibility \parencite{dableheath2025corrigibilityalignmentmultiagent,carey2023humancontroldefinitionsalgorithms}, scalable oversight \parencite{kenton2024scalableoversightweakllms, irving2018aisafetydebate}, Constitutional AI 
\parencite{bai2022constitutionalaiharmlessnessai}, and Deliberative Alignment \parencite{guan2025deliberativealignmentreasoningenables}. However, it is unclear whether and to what extent any of these approaches or other future approaches will be successful and/or sufficient. Second, even if we assume that scientific ingenuity would solve misalignment in time, pure malfunction presents a potentially significant, residual risk of an AI system behaving in entirely unpredictable and uncontrollable ways.\footnote{Threats catalyzed by misalignment and/or malfunction are compounded by an AI system’s capabilities and propensities, the deployment context the AI system is in, and its affordances and permissions.
} We explore the implications next.
\phantomsection
\subsection*{3.C Implications of Living with a State of Vulnerability}\label{sov_implication}

Previously, we laid out reasons as to why we may arrive at a societal state of vulnerability (\textit{see} \hyperref[sov_pathway]{Section 3.A}) and what we propose to be the catalyst for LoC (\textit{see} \hyperref[catalyst]{Section 3.B}). Now, we build on these reflections and present a high-level logical sequence that enables us to speculate as to what is likely to happen once we find ourselves in a state of vulnerability to LoC. In other words, we now concentrate on what we believe could plausibly happen once a sufficiently capable AI system has acquired (through humans or independently) or could independently acquire sufficient access to resources, affordances, and permissions (or means to acquire further access) and sufficient capabilities to cause LoC.

\begin{figure}[h!]
    \centering
    \captionsetup{justification=justified}
    \includegraphics[width=0.8\linewidth]{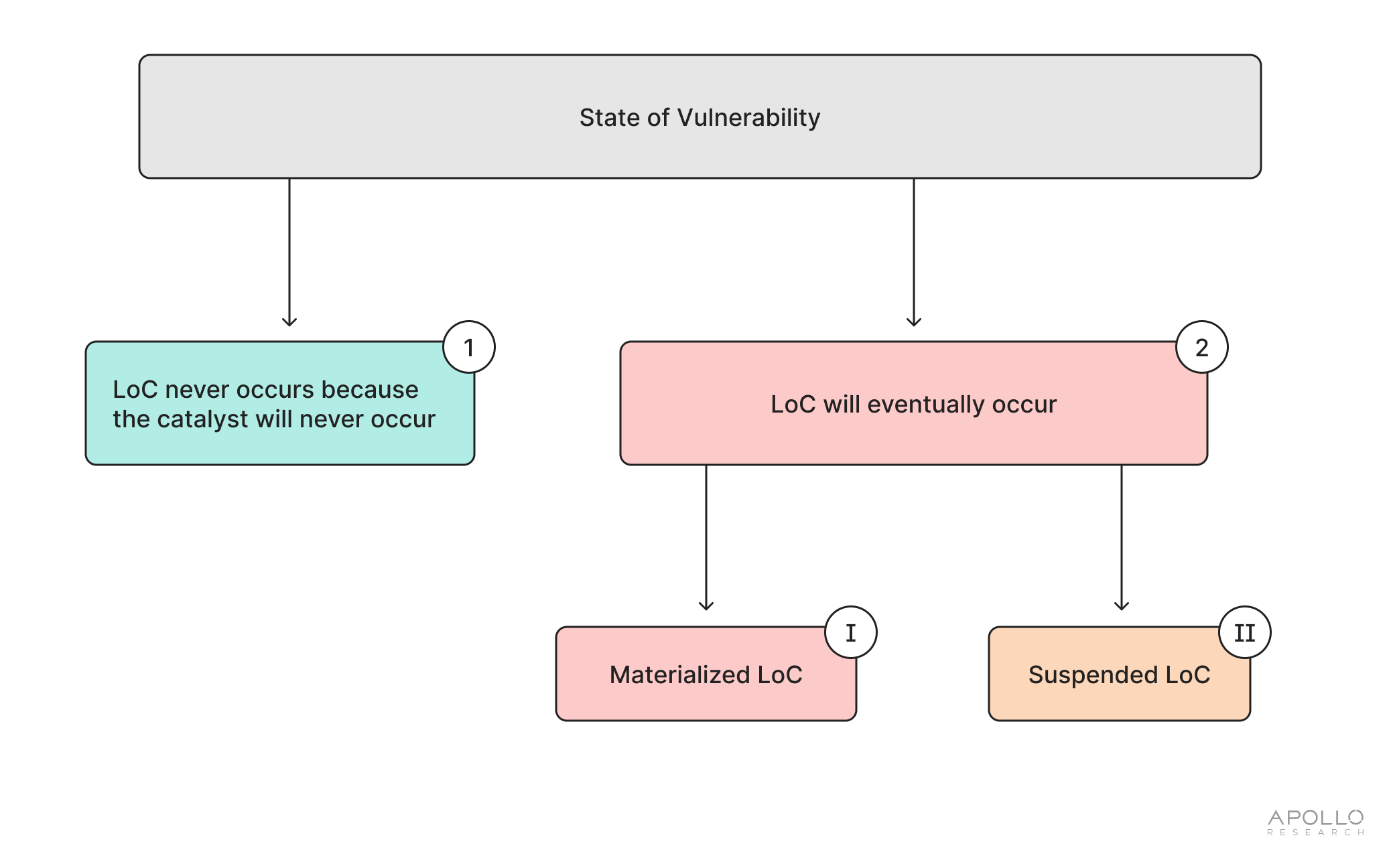}
    \caption{A simple diagram illustrating pathways that may arise 
after society has found itself living with a state of vulnerability.}
    \label{figure6:SoV_simple}
\end{figure}

Simply put, there are only two possible outcomes once we are in a state of vulnerability (\textit{see}~\autoref{figure6:SoV_simple}). \textbf{Namely, a state of vulnerability either (1) does not lead to LoC, or (2) leads to LoC}.

\textbf{In order for a state of vulnerability to (1) not lead to LoC, we would have to assume that the AI system in question lacks any catalyst towards LoC}. Concretely, this means that \textbf{the AI system is neither affected by misalignment nor pure malfunction}. Going back to our bomb metaphor, this means that, even if all other conditions for an explosion exist, the bomb can never be ignited. 

While it is theoretically possible that a ‘completely safe world’ exists (outcome (1)), where LoC would never materialize despite an existing state of vulnerability, we believe this outcome is highly improbable. Such a world would necessitate that neither of the components of the catalyst for LoC (i.e., misalignment and pure malfunctions) ever come to fruition. \textbf{In other words, even if the alignment problem were solved, AI systems would be required to function perfectly for the entirety of their service periods, absent any pure malfunction}. Given that malfunctions are often unknown unknowns until they occur and therefore their nature and future avoidance can only be known post facto, the chance that there will never be any pure malfunction throughout a given AI system’s service period is slim. Moreover and importantly, even if we could, in fact, end up in a ‘safe world’ scenario, \textbf{it is unlikely that we would be able to reliably prove this ahead of time}. While we may be able to assure that misalignment will not occur, one would be required to make an affirmative case that pure malfunction is also impossible. This affirmative case appears to be extraordinarily difficult to establish, given the nature of pure malfunctions. Applying the precautionary principle to these uncertainties, we should therefore reasonably assume that we are in a world in which the state of vulnerability does eventually lead to LoC, i.e., outcome (2) and not outcome (1). 

Outcome (2) captures the ‘unsafe world.’ In other words, it captures a situation in which a state of vulnerability will eventually be ignited through a catalyst, and a LoC threat will materialize. In this unsafe world, LoC outcome could either: (i) \textbf{immediately materialize} (‘materialized LoC’); or (ii) \textbf{materialize at some future point }(‘suspended LoC’). 

\textbf{If (2.i) occurs, then a LoC outcome has materialized. }Since we are referring to LoC to capture Bounded LoC and Strict LoC (\textit{see} \hyperref[reflections_findings]{Section 1.A.2.b}), this materialization implies that society finds itself faced with an outcome in either of these two categories. Depending on the severity of the Bounded LoC, which will be hard to predict ex ante, society may be willing to incur substantial costs to avoid such an outcome. Certainly, society will wish to avoid the materialization of Strict LoC, which poses an irreversible situation with no scope left for action. \textbf{In both instances, branch (2.i) implies that a LoC outcome has occurred, and therefore, this is the outcome to avoid at high or all costs.}

We call \textbf{branch (2.ii) ‘suspended LoC’ since it presents a theoretical state where materialization of a LoC outcome will occur at a future point,}\textbf{ but hasn’t yet; it is, therefore, suspended}.\footnote{ The occurrence of this LoC threat could be quick or devolve over time. In both cases, LoC will occur at some point in the future.} For instance, an AI system might be accumulating resources, but no LoC outcome has occurred yet. In other words, this situation accounts for the ‘creeping problem’ that LoC can abstractly pose in worlds where it does not instantaneously materialize. Alternatively, LoC could be suspended because the safeguards placed upon the AI system are sufficient to contain it. Importantly, in worlds where potential LoC is suspended, it is plausible that \textbf{we will lack information about both whether and why} \textbf{it is suspended}. This uncertainty is likely to pose challenges to ensuring that a given AI system does not progress into materialized LoC.

\begin{figure}[h!]
    \centering
    \captionsetup{justification=justified}
    \includegraphics[width=0.9\linewidth]{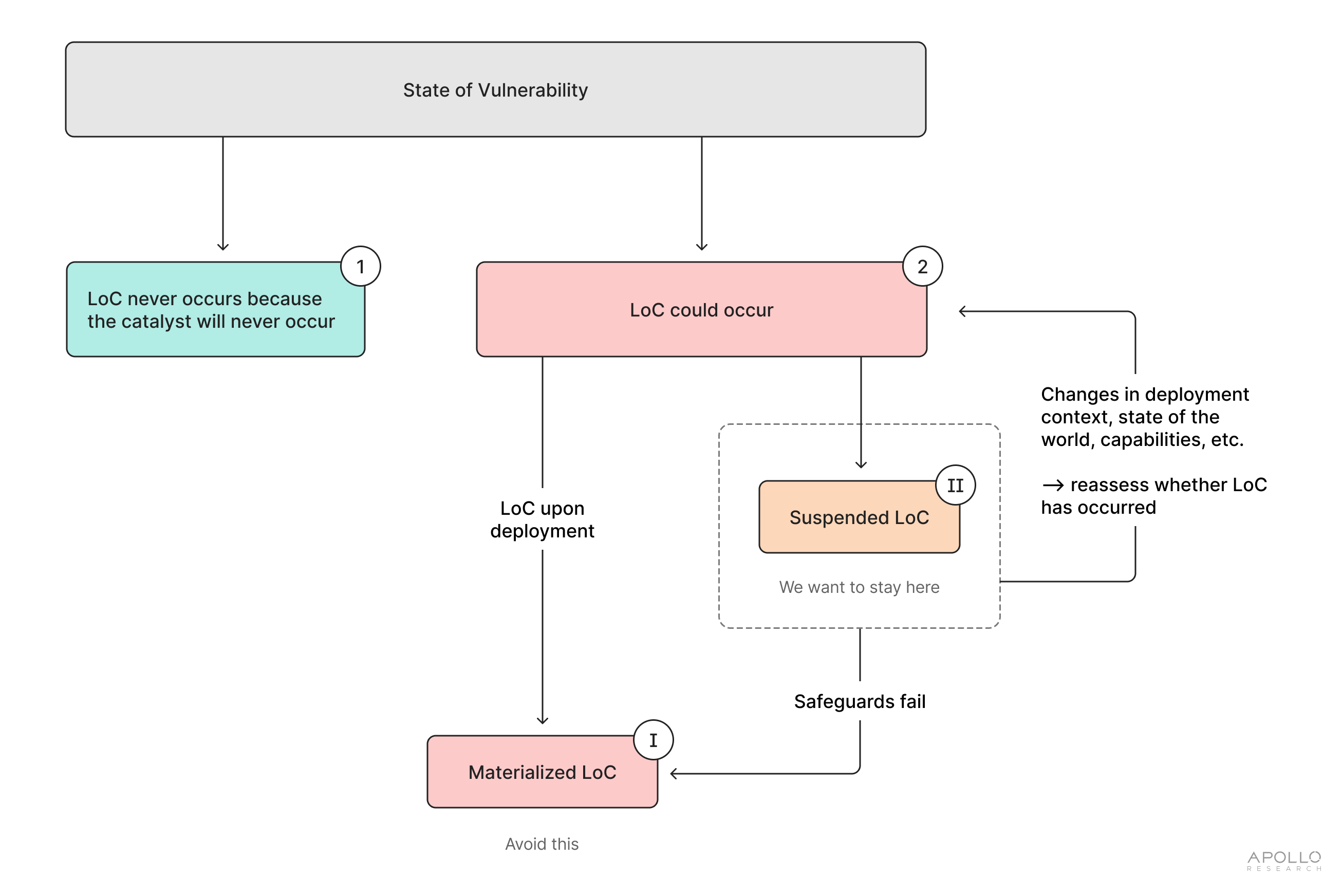}
    \caption{An illustration demonstrating the changing nature of an AI 
system’s risk profile that one must account for, and the corresponding
replay of the first branching.}
    \label{figure7:SoV_complex}
\end{figure}

In this respect, we observe that continuously upholding suspension and making it feasible to live in a perennial state of vulnerability will be challenging. Not least because intrinsic factors (capabilities and propensities) and extrinsic factors (deployment context, affordances, and permissions) are likely to differ between distinct AI systems and are subject to change throughout a given AI system’s service period. With every change in these intrinsic and extrinsic factors, the risk profile changes, and the setup could lead to materialized LoC (2.i). In other words, each time the context changes, we have to assume that society is replaying the initial branching (\textit{see}~\autoref{figure7:SoV_complex}).

Given the above, we therefore conclude that once we live with a state of vulnerability, unless there can be a guarantee that the probability of a catalyst is 0 and that society will therefore find itself in a world in which LoC never occurs, \textbf{every deployment is a roll of the dice}.
\phantomsection
\subsection*{3.D Reflections}\label{sov_reflections}

We have previously elaborated on why it is likely that society will reach a state of vulnerability and why it is likely that such a state will devolve into LoC. We conclude this chapter by emphasizing that, although the odds may seem unfavorable, these outcomes are not yet set in stone and that action is still possible. As part of this, we put forward a number of reflections below. 

First and foremost, the most robust intervention today would be as follows:

\begin{itemize}
    \item Society should \textbf{work toward avoiding a state of vulnerability} and \textbf{reducing potential catalysts }(misalignment and malfunction). In other words, deployers and policymakers could choose to ensure that highly advanced future AI systems are not integrated in manners and deployment contexts that pose a higher risk of LoC (\textit{see} \hyperref[chapter-2]{Chapter 2}). As the incentives to deploy more capable and autonomous AI systems in higher-stakes contexts increase, developers, deployers, and policymakers must lucidly prepare for the implications of living with a state of vulnerability, including the uncertainty of fully resolving misalignment, the unknown consequences of pure malfunctions, and the possible outcomes of a materialized LoC scenario of various scales (\textit{see} \hyperref[chapter-1]{Chapter 1}).
\end{itemize}

While we do not know when the ‘threshold’ toward a state of vulnerability will be crossed, we expect that, at least theoretically, it could be feasible to maintain a state of suspended LoC over time, assuming we have adequate and robust defence-in-depth mechanisms, spanning state-of-the-art threat modeling, testing, oversight, and control mechanisms in place. 

We therefore suggest that, should a state of vulnerability be reached in the future, then:

\begin{itemize}
    \item Society should \textbf{attempt to maintain a perennial state of suspension. }One can never be certain that a state of vulnerability will not lead to a LoC outcome. A sensible course of action may therefore be to act as if one lives in a world where LoC could materialize, but has not yet––in other words, in a world of suspension.
\end{itemize}

A comprehensive treatment of what is required to maintain suspension of LoC is outside the scope of this report. However, we indicate that an adequate defense-in-depth approach should likely, at a minimum, consider the following two categories:

\begin{itemize}
    \item \textbf{Governance interventions }including, among others: concrete threat modeling \parencite{nsa_deploying_ai_2024, OWASP_Threat_Modeling_Community_Page}; policies that describe acceptable deployments \parencite{Microsoft2025_BuiltInPolicyModelDeployment}; and wide-reaching, easy-to-enact emergency response plans \parencite{stix2025aicloseddoorsprimer, RAND_Preparedness, RAND_LOC_Response_Planning, RAND_Countering_Rogue}.
    \item \textbf{Technical interventions} including, among others: rigorous pre-deployment testing suites in accordance with threat models for the deployment context; control measures that constrain an AI system’s effect on the world around the AI system (\parencite{greenblatt2024aicontrolimprovingsafety,korbak2025sketchaicontrolsafety}; as well as stringent human and AI-enabled monitoring \parencite{ncsc_secure_ai_2023}.\footnote{ We want to be explicit that: (i) the spectrum of these interventions may change and that interventions should adapt to the most robust future mechanisms available; that (ii) these interventions are not comprehensive; and that (iii) these intervention categories should not be misconstrued as containing independently sufficient features but instead should always be considered holistically.}
\end{itemize}

%% file: content/Conclusion.tex
\phantomsection
\section*{Conclusion}\label{conclusion}
In this research report, we provided \textbf{a novel} \textbf{conceptualization, taxonomy, and future implications of LoC as a threat. }Our work was motivated by recent increases in AI system capabilities and rising attention to the topic in policy, industry, and legal discourse. 

In Chapter 1, we set out to shed light on \textbf{what LoC is}. We arrived at our LoC taxonomy by reviewing relevant definitions in AI literature and in other safety-critical sectors, including defense, aviation, and nuclear. We found that neither set of existing definitions supports one common, clear definition of LoC that decision- and policymakers could leverage today. Subsequently, we conducted a comprehensive literature review and developed a methodology to assess and categorize existing LoC scenarios, ultimately plotting them on an experimental graph for comparison. Our graph enabled us to extrapolate \textbf{a novel taxonomy for conceptualizing LoC: Deviation, Bounded LoC, and Strict LoC}. 

We believe that this taxonomy will be helpful in structuring discussions around LoC, as each category entails different outcomes in terms of permanence and severity, and may potentially require different mitigation strategies and actionable levers. Moreover, our methodology sheds light on which category is most commonly and concretely referenced in scholarly works: Bounded LoC.

In Chapter 2, we reflected on \textbf{what can be done to mitigate LoC threats today}. We proposed that in the absence of a consensus on capabilities and thresholds to accurately assess and capture LoC risk, decision- and policymakers should focus on actionable, simple steps that can be taken today. We therefore proposed a complementary approach that can offer clear levers, while largely sidestepping existing uncertainties. Specifically, instead of focusing on AI systems’ intrinsic factors (i.e., capabilities and propensities), we proposed to focus on extrinsic factors that can raise the overall risk of LoC occurring. In doing so, we put forward a framework inspired by the research presented in Chapter 1, focusing on assessing and limiting the \textbf{deployment context, affordances, and permissions of an AI system (the ‘DAP framework’)}.

In Chapter 3, we focused on the \textbf{future implications of LoC threats,} in light of the assumed rise in AI capabilities and increased strategic and economic competition, which could undermine the DAP framework. We claim that under these conditions, \textbf{society would eventually find itself living in a }‘\textbf{state of vulnerability},’ which denotes a state in which a sufficiently capable future AI system has acquired (through humans or independently) or could independently acquire sufficient access to resources, affordances, and permissions (or means to acquire further access) and sufficient capabilities to cause LoC when a catalyst materializes. Once the catalyst is triggered, we speculate on multiple pathways that future could take and propose that it is implausible for society not to eventually face a LoC outcome; therefore, preemptive control and mechanisms should be developed and implemented early on.

%% file: content/Appendix.tex
\section*{Appendix}\label{appendix}
\subsection*{Appendix 1: Future Research Directions}\label{appendix1}
In order to encourage future research beyond the targeted scope of this report, we outline below some research directions that could shed light on the open questions we encountered during our investigation. 

\begin{enumerate}[nosep]
    \item Research on LoC scenarios in which severity and persistence may not strongly correlate (e.g., scenarios that are high-severity but low-persistence or low-severity but high-persistence). 
    \item Analysis of which types of Deviation are likely to be warning shots for more significant LoC incidents, to what extent we can extrapolate from them, and what interventions they should trigger.
    \item Relatedly, research on which indicators can be used to determine whether an incident deemed as a Deviation is likely to be a warning shot for more severe events that might culminate in Bounded LoC or Strict LoC. 
    \item Quantifying the current state of LoC risk and assessing the likelihood of potential LoC scenarios. 
    \begin{itemize}[nosep]
        \item Which quantitative statements can be made about the LoC risk stemming from current systems? 
        \item How likely are the scenarios from the literature (or analogous ones) to materialize? 
    \end{itemize}
    \item Which capabilities and propensities most increase the risk of LoC?
    \item Which combinations of capabilities, propensities (and DAPs) present compounding risks, i.e., the resulting risk is greater than the sum of its parts?
    \item Which sequences of events are likely to lead to LoC? \footnote{Answers to this research question could allow more targeted mitigations and more detailed threat modeling.}
    \item At which thresholds of capabilities and propensities should further safeguards to prevent LoC be implemented?
    \item Which safeguards against LoC are appropriate and effective?
    \item How can we reduce the likelihood of a catalyst (misalignment/malfunction) occurring? 
    \item How should governance and technical interventions be implemented for a given deployment scenario? \footnote{This could include creating best-practice risk management procedures for certain industries (e.g., critical infrastructure).}
    \item Specifically, which affordances and permissions are (and are not) strictly necessary for various use cases in high-stakes deployment contexts?
    \item How should disempowerment risks from AI be quantified, in cases in which disempowerment does not necessarily imply human extinction? 
    \item Which sets of metrics and parameters should be adopted to classify deployment contexts as high-stakes?
\end{enumerate}

\subsection*{Appendix 2: Additional Context on the Methodology for our Taxonomy of LoC}
In this Appendix, we provide additional context on the assumptions and methodology underpinning our taxonomy of LoC described in Chapter 1 and visually represented in~\autoref{figure2:the_graph}. Specifically, we provide additional context on:
\begin{enumerate}[label=\textbf{2.\arabic*}]
\item The works we reviewed for our literature review as described in 1.A.1. 
\item Building on our literature review, the two criteria, and their respective sub-criteria, that we adopted for deeming a LoC scenario, as ‘concrete’ or ‘not concrete’ as described in 1.A.2.a.
\item[\textbf{2.2.1}] Additional details on our concrete LoC scenarios, including on their economic impact estimates. 
\end{enumerate}

\subsubsection*{2.1 Literature Reviewed}\label{appendix2.1}
We reviewed a total of 130 works, including academic papers, research reports, publications from various think tanks, and government publications, in order to cast our net wide. The works reviewed are as follows: 

Kokotajlo, Daniel, Scott Alexander, Thomas Larsen, Eli Lifland, and Romeo Dean. 2025. ``AI 2027: A
research-backed AI scenario forecast.”  \url{https://ai-2027.com/}.

Dung, Leonard. 2025. ``The argument for near-term human disempowerment through AI.” \url{https://link.springer.com/article/10.1007/s00146-024-01930-2}.

Carlsmith, Joseph. 2024. ``Is Power-Seeking AI an Existential Risk?" \url{https://arxiv.org/abs/2206.13353}. 
Critch, Andrew, and Stuart Russell. 2023. ``TASRA: a Taxonomy and Analysis of Societal-Scale Risks from
AI". \url{https://arxiv.org/abs/2306.06924}.

Barnett, Peter, and Jeremy Gillen. 2024. ``Without Fundamental Advances, Misalignment and Catastrophe
Are the Default Outcomes of Training Powerful AI". \url{https://intelligence.org/wp-content/uploads/2024/12/Misalignment and Catastrophe.pdf}.

Kalra, Nidhi, and Benjamin Boudreaux. 2025. ``Not Just Superintelligence: The Many Risks of Near-Future
AGI,” \url{https://geopoliticsagi.substack.com/p/not-just-superintelligence-the-many}.

Bengio, Yoshua, Michael Cohen, Damiano Fornasiere, Joumana Ghosn, Pietro Greiner, et al. 2025. ``Superintelligent Agents Pose Catastrophic Risks: Can Scientist AI
Offer a Safer Path?". \url{https://arxiv.org/abs/2502.15657}.

Shah, Rohin, Vikrant Varma, Ramana Kumar, Mary Phuong, Victoria Krakovna, et al. 2022. ``Goal Misgeneralization: Why Correct Specifications Aren’t Enough For Correct Goals".
\url{https://arxiv.org/abs/2210.01790}.

Sotala, Kaj. 2018. ``Disjunctive Scenarios of Catastrophic AI Risk". \url{https://kajsotala.fi/assets/2018/12/Disjunctivescenarios.pdf}.

Kulveit, Jan, Raymond Douglas, Nora Ammann, Deger Turan, David Krueger, et al. 2025.
``Gradual Disempowerment: Systemic Existential Risks from Incremental AI Development". \url{https://arxiv.org/abs/2501.16946}.

Clarke, Sam, Jess Whittlestone. 2022. ``A Survey of the Potential Long-term Impacts of AI." \url{https://dl.acm.org/doi/pdf/10.1145/3514094.3534131}.

Hendrycks, Dan, Eric Schmidt, Alexandr Wang. 2025. ``Superintelligence Strategy: Expert Version." \url{https://arxiv.org/pdf/2503.05628}. 

Hendrycks, Dan, Mantas Mazeika, and Thomas Woodside. 2023. ``An Overview of Catastrophic AI Risks". \url{https://arxiv.org/abs/2306.12001}.

Ngo, Richard, Lawrence Chan, Sören Mindermann. 2025. ``The Alignment Problem from a Deep Learning Perspective". \url{https://arxiv.org/pdf/2209.00626}.

Pavel, Barry, Ivana Ke, Gregory Smith, Sophia Brown-Heidenreich, Lea Sabbag, et al. 2025. ``How Artificial General Intelligence Could Affect the Rise and Fall of Nations:
Visions for Potential AGI Futures." \url{https://doi.org/10.7249/RRA3034-2}.

Özcan, Bengüsu, Daan Juijn, Jakob Graabak, Sam Bogerd. 2025. ``Advanced AI: Possible futures". \url{https://cfg.eu/advanced-ai-possible-futures/#take-off}.

Bengio, Yoshua. 2024. ``Government Interventions to Avert Future Catastrophic AI Risks". \url{https://hdsr.mitpress.mit.edu/pub/w974bwb0/release/2}.

Hendrycks, Dan. 2023. ``Natural Selection Favors AIs over Humans". \url{https://arxiv.org/abs/2303.16200}.

Shah, Rohin, Alex Irpan, Alexander Matt Turner, Anna Wang, Arthur Conmy et al. 2025. ``An Approach to Technical AGI Safety and Security". \url{https://arxiv.org/pdf/2504.01849}.

Raman, Deepika, Nada Madkour, Evan R. Murphy, Krystal Jackson, Jessica Newman. 2025. ``Intolerable Risk Threshold Recommendations for Artificial Intelligence" \url{https://www.aigl.blog/content/files/2025/04/Intolerable-Risk-Threshold-Recommendations-for-Artificial-Intelligence.pdf}.

Somani, Elika, Anjay Friedman, Henry Wu, Marianne Lu, Christopher Byrd, et al. 2025. ``Strengthening Emergency Preparedness and Response for AI Loss of Control Incidents." \url{https://doi.org/10.7249/RRA3847-1}.

Wasil, Akash R., Everett Smith, Corin Katzke, Justin Bullock. 2024. ``AI Emergency Preparedness: Examining the federal government’s ability to detect and respond to AI-related national security threats" \url{https://arxiv.org/pdf/2407.17347}.

Salmon, Paul M., Tony Carden, Peter A. Hancock. 2020. ``Putting the humanity into inhuman systems: How human factors and ergonomics can be used to manage the risks associated with artificial general intelligence". \url{https://doi.org/10.1002/hfm.20883}.

Department for Science, Innovation \& Technology (DSIT). 2025. ``Future risks of frontier AI (Annex A)". \url{https://www.gov.uk/government/publications/frontier-ai-capabilities-and-risks-discussion-paper/future-risks-of-frontier-ai-annex-a}.

Bernardi, Jamie, Gabriel Mukobi, Hilary Greaves, Lennart Heim, Markus Anderljung. 2025. ``Societal Adaptation to Advanced AI". \url{https://arxiv.org/pdf/2405.10295}. 

Drexel, Bill, Caleb Withers. 2024. ``Catalyzing Crisis: A Primer on Artificial Intelligence, Catastrophes, and National Security." \url{https://s3.us-east-1.amazonaws.com/files.cnas.org/documents/Catastrophic-AI_TECH-2024_Final.pdf}.

Bengio, Yoshua, Sören Mindermann, Daniel Privitera, Tamay Besiroglu, Rishi Bommasani, et al. 2025. ``International AI Safety Report 2025." \url{https://internationalaisafetyreport.org/publication/international-ai-safety-report-2025}.

Cass-Beggs, Duncan, Stephen Clare, Dawn Dimowo, Zaheed Kara. 2024. ``Framework Convention on Global AI Challenges: Accelerating International Cooperation to Ensure Beneficial, Safe and Inclusive AI." \url{https://www.cigionline.org/static/documents/AI-challenges_OW6rTMD.pdf}.

Bengio, Yoshua, Geoffrey Hinton, Andrew Yao, Dawn Song, Pieter Abbeel, et al. 2024. ``Managing Extreme AI Risks amid Rapid Progress." \url{https://arxiv.org/abs/2310.17688}.

Barnett, Peter, Aaron Scher, David Abecassis. 2025. ``Technical Requirements for Halting Dangerous AI Activities." \url{https://arxiv.org/abs/2507.09801}.

Park, Peter S., Simon Goldstein, Aidan O'Gara, Michael Chen, Dan Hendrycks. 2023. ``AI Deception: A Survey of Examples, Risks, and Potential Solutions." \url{https://arxiv.org/abs/2308.14752}.

Bales, Adam, William D'Alessandro, Cameron Domenico Kirk-Giannini. 2024. ``Artificial Intelligence: Arguments for Catastrophic Risk." \url{https://arxiv.org/abs/2401.15487}.

Motwani, Sumeet Ramesh, Mikhail Baranchuk, Martin Strohmeier, Vijay Bolina, Philip H.S. Torr, et al. 2025. ``Secret Collusion among AI Agents: Multi-Agent Deception via Steganography." \url{https://arxiv.org/abs/2402.07510}.

Barnett, Peter, Aaron Scher. 2025. ``AI Governance to Avoid Extinction: The Strategic Landscape and Actionable Research Questions." \url{https://arxiv.org/abs/2505.04592}.

Jones, Charles I. 2024. ``The AI Dilemma: Growth versus Existential Risk." \url{https://web.stanford.edu/~chadj/existentialrisk.pdf}.

Swoboda, Torben, Risto Uuk, Lode Lauwaert, Andrew P. Rebera, Ann-Katrien Oimann, et al. 2025. ``Examining Popular Arguments Against AI Existential Risk: A Philosophical Analysis." \url{https://arxiv.org/pdf/2501.04064}.

Miotti, Andrea. 2025. ``Taking control: Policies to address extinction risks from advanced AI." \url{https://arxiv.org/pdf/2310.20563}.

Phuong, Mary, Roland S. Zimmermann, Ziyue Wang, David Lindner, Victoria Krakovna, et al. 2025. ``Evaluating Frontier Models for Stealth and Situational Awareness." \url{https://arxiv.org/pdf/2505.01420}.

Bucknall, Benjamin S., Shiri Dori-Hacohen. 2022. ``Current and Near-Term AI as a Potential Existential Risk Factor." \url{https://arxiv.org/abs/2209.10604}.

Russell, Stuart. 2022. ``Artificial Intelligence and the Problem of Control." \url{https://link.springer.com/chapter/10.1007/978-3-030-86144-5_3}.

Li, Xiaojian, Haoyuan Shi, Rongwu Xu, Wei Xu. 2025. ``AI Awareness." \url{https://arxiv.org/pdf/2504.20084}.

Dung, Leonard. 2025. ``Evaluating approaches for reducing catastrophic risks from AI." \url{https://link.springer.com/article/10.1007/s43681-024-00475-w}.

Kingsley, Jeremy, Syedah Ailia Haider, Shreyansh Jain, Adam Green, Jan Copeman, et al. 2024. ``AI landscapes: Exploring future scenarios of AI through to 2030." \url{https://securesustain.org/wp-content/uploads/2024/05/AI-Landscapes-Exploring-Future-Scenarios-of-AI-Through-to-2030.pdf}.

Lynch, Aengus, Benjamin Wright, Caleb Larson, Stuart J. Ritchie, Kevin K. Troy, et al. 2025. ``Agentic Misalignment: How LLMs could be insider threats." \url{https://www.anthropic.com/research/agentic-misalignment}.

OSTP. 2025. ``America’s AI Action Plan". \url{https://www.whitehouse.gov/wp-content/uploads/2025/07/Americas-AI-Action-Plan.pdf}.
        
Benavoli, Alessio, Alessandro Facchini, Marco Zaffalon. 2025. ``The AI off-switch problem as a signalling game: bounded rationality and incomparability." \url{https://arxiv.org/abs/2502.06403}.

Hastings-Woodhouse, Sarah. 2024. ``Could we switch off a dangerous AI?" \url{https://futureoflife.org/ai/could-we-switch-off-a-dangerous-ai/}.

Schmitz, Chris, Jonathan Rystrøm, Jan Batzner. 2025. ``Oversight Structures for Agentic AI in Public-Sector Organizations." \url{https://arxiv.org/abs/2506.04836}.

Humble, Kristian. 2024. ``War, Artificial Intelligence, and the Future of Conflict." \url{https://gjia.georgetown.edu/2024/07/12/war-artificial-intelligence-and-the-future-of-conflict/}.

Pan, Alexander, Kush Bhatia, Jacob Steinhardt. 2022. ``The Effects of Reward Misspecification: Mapping and Mitigating Misaligned Models." \url{https://arxiv.org/abs/2201.03544}.

Hendrycks, Dan, Nicholas Carlini, John Schulman, Jacob Steinhardt. 2021. ``Unsolved Problems in ML Safety." \url{https://arxiv.org/pdf/2109.13916}.

Omohundro, Stephen M. 2008. ``The Basic AI Drives." \url{https://selfawaresystems.com/wp-content/uploads/2008/01/ai_drives_final.pdf}.

Wei, Jason, Yi Tay, Rishi Bommasani, Colin Raffel, Barret Zoph, et al. 2022. ``Emergent Abilities of Large Language Models." \url{https://openreview.net/pdf?id=yzkSU5zdwD}.

Meinke, Alexander, Bronson Schoen, Jérémy Scheurer, Mikita Balesni, Rusheb Shah, et al. 2025. ``Frontier Models are Capable of In-context Scheming." \url{https://arxiv.org/abs/2412.04984}.

Rudner, Tim G. J., Helen Toner. 2021. ``Key Concepts in AI Safety: Specification in Machine Learning." \url{https://cset.georgetown.edu/wp-content/uploads/Key-Concepts-in-AI-Safety-Specification-in-Machine-Learning.pdf}.

Langosco di Langosco, Lauro, Jack Koch, Lee D. Sharkey, Jacob Pfau, David Krueger. 2022. ``Goal Misgeneralization in Deep Reinforcement Learning." \url{https://proceedings.mlr.press/v162/langosco22a/langosco22a.pdf}.

Fang, Richard, Rohan Bindu, Akul Gupta, Daniel Kang. 2024. ``LLM Agents can Autonomously Exploit One-day Vulnerabilities." \url{https://arxiv.org/pdf/2404.08144}.

Dalrymple, David, Joar Skalse, Yoshua Bengio, Stuart Russell, Max Tegmark, et al. 2024. ``Towards Guaranteed Safe AI: A Framework for Ensuring Robust and Reliable AI Systems." \url{https://arxiv.org/abs/2405.06624}.

Fang, Richard, Rohan Bindu, Akul Gupta, Qiusi Zhan, Daniel Kang. 2024. ``LLM Agents can Autonomously Hack Websites." \url{https://arxiv.org/abs/2402.06664}.

Christiano, Paul F., Jan Leike, Tom B. Brown, Miljan Martic, Shane Legg, et al. 2017. ``Deep Reinforcement Learning from Human Preferences." \url{https://dl.acm.org/doi/pdf/10.5555/3294996.3295184}.

Bai, Yuntao, Andy Jones, Kamal Ndousse, Amanda Askell, Anna Chen, et al. 2022. ``Training a Helpful and Harmless Assistant with Reinforcement Learning from Human Feedback." \url{https://arxiv.org/abs/2204.05862}.

Armstrong, Stuart, Xavier O'Rorke. 2018. ``Good and Safe Uses of AI Oracles." \url{https://arxiv.org/abs/1711.05541}.

Narayanan, Arvind, Sayash Kapoor. 2025. ``AI as Normal Technology." \url{https://knightcolumbia.org/content/ai-as-normal-technology}.

Grace, Katja, Harlan Stewart, Julia Fabienne Sandkühler, Stephen Thomas, Ben Weinstein-Raun, et al. 2025. ``Thousands of AI Authors on the Future of AI." \url{https://arxiv.org/pdf/2401.02843}.

Bai, Yuntao, Saurav Kadavath, Sandipan Kundu, Amanda Askell, Jackson Kernion, et al. 2022. ``Constitutional AI: Harmlessness from AI Feedback." \url{https://arxiv.org/abs/2212.08073}.

Turner, Alexander Matt, Logan Smith, Rohin Shah, Andrew Critch, Prasad Tadepalli. 2021. ``Optimal Policies Tend to Seek Power." \url{https://openreview.net/pdf?id=l7-DBWawSZH}.

Matheny, Jason. 2023. ``Artificial Intelligence: Challenges and Opportunities for the Department of Defense." \url{https://www.rand.org/content/dam/rand/pubs/testimonies/CTA2700/CTA2723-1/RAND_CTA2723-1.pdf}.

Taylor, Jessica, Eliezer Yudkowsky, Patrick LaVictoire, Andrew Critch. 2016. ``Alignment for Advanced Machine Learning Systems." \url{https://intelligence.org/files/AlignmentMachineLearning.pdf}.

Askell, Amanda, Miles Brundage, Gillian Hadfield. 2019. ``The Role of Cooperation in Responsible AI Development." \url{https://arxiv.org/abs/1907.04534}.

Kasirzadeh, Atoosa. 2025. ``Two types of AI existential risk: decisive and accumulative." \url{https://link.springer.com/article/10.1007/s11098-025-02301-3}.

Clark, Jack, Dario Amodei. 2016. ``Faulty reward functions in the wild." \url{https://openai.com/index/faulty-reward-functions/}.

De Marzo, Giordano, Claudio Castellano, David Garcia. 2025. ``AI agents can coordinate beyond human scale." \url{https://arxiv.org/abs/2409.02822}.

Milli, Smitha, Dylan Hadfield-Menell, Anca Dragan, Stuart Russell. 2017. ``Should Robots be Obedient?" \url{https://arxiv.org/abs/1705.09990}.

Armstrong, Stuart, Benjamin Levinstein. 2017. ``Low Impact Artificial Intelligences." \url{https://arxiv.org/abs/1705.10720}.

Boling, Bryan, Benjamin Boudreaux, Alexis A. Blanc, Christy Foran, Edward Geist. 2022. ``Emerging Technology Beyond 2035: Scenario-Based Technology Assessment for Future Military Contingencies." \url{https://www.rand.org/pubs/research_reports/RRA1564-1.html}.

Bullock, Justin, Corin Katzke, Zershaaneh Qureshi, David Kristoffersson. 2024. ``AI Clarity: An Initial Research Agenda." \url{https://www.convergenceanalysis.org/publications/ai-clarity-an-initial-research-agenda}.

Qureshi, Zershaaneh. 2025. ``Pathways to short TAI timelines." \url{https://www.convergenceanalysis.org/research/pathways-to-short-tai-timelines}.

Katzke, Corin, Justin Bullock. 2024. ``AI governance needs a theory of victory." \url{https://www.convergenceanalysis.org/publications/ai-governance-needs-a-theory-of-victory}.

Katzke, Corin. 2024. ``Investigating the role of agency in AI x-risk." \url{https://www.convergenceanalysis.org/publications/investigating-the-role-of-agency-in-ai-x-risk}.

Katzke, Corin. 2024. ``Scenario planning for AI x-risk." \url{https://www.convergenceanalysis.org/publications/scenario-planning-for-ai-x-risk}.

Bullock, Justin, Elliot McKernon. 2024. ``Transformative AI and Scenario Planning for AI X-risk." 
\url{https://www.convergenceanalysis.org/publications/transformative-ai-and-scenario-planning-for-ai-x-risk}.

Dung, Leonard. 2025. ``Understanding Artificial Agency." \url{https://academic.oup.com/pq/article-abstract/75/2/450/7601099?redirectedFrom=fulltext&login=false}.

Dung, Leonard. 2024. ``Is superintelligence necessarily moral?" \url{https://academic.oup.com/analysis/article-abstract/84/4/730/7774058?redirectedFrom=fulltext&login=false}.

Slattery, Peter, Alexander K. Saeri, Emily A. C. Grundy, Jess Graham, Michael Noetel. 2024. ``The AI Risk Repository: A Comprehensive Meta-Review, Database, and Taxonomy of Risks From Artificial Intelligence." \url{https://arxiv.org/abs/2408.12622}.

Hubinger, Evan, Chris van Merwijk, Vladimir Mikulik, Joar Skalse, Scott Garrabrant. 2021. ``Risks from Learned Optimization in Advanced Machine Learning Systems." \url{https://arxiv.org/abs/1906.01820v3}.

Gabriel, Iason, Arianna Manzini, Geoff Keeling, Lisa Anne Hendricks, Verena Rieser. 2024. ``The Ethics of Advanced AI Assistants." \url{https://storage.googleapis.com/deepmind-media/DeepMind.com/Blog/ethics-of-advanced-ai-assistants/the-ethics-of-advanced-ai-assistants-2024-i.pdf}.

Ji, Jiaming, Tianyi Qiu, Boyuan Chen, Borong Zhang, Hantao Lou. 2023. ``AI Alignment: A Comprehensive Survey." \url{https://arxiv.org/pdf/2310.19852}.

AI Verify Foundation. 2023. ``Cataloguing LLM Evaluations." \url{https://aiverifyfoundation.sg/downloads/Cataloguing_LLM_Evaluations.pdf}.

Shevlane, Toby, Sebastian Farquhar, Ben Garfinkel, Mary Phuong, Jess Whittlestone. 2023. ``Model evaluation for extreme risks." \url{https://arxiv.org/pdf/2305.15324}.

Yudkowsky, Eliezer. 2008. ``Artificial Intelligence as a Positive and Negative Factor in Global Risk." \url{https://intelligence.org/files/AIPosNegFactor.pdf}.

Bradley, Patrick. 2020. ``Risk management standards and the active management of malicious intent in artificial superintelligence." \url{https://link.springer.com/article/10.1007/s00146-019-00890-2}.

Sotala, Kaj, Roman V. Yampolskiy. 2014. ``Responses to Catastrophic AGI Risk: A Survey." \url{https://iopscience.iop.org/article/10.1088/0031-8949/90/1/018001}.

Barrett, Anthony M., Seth D. Baum. 2017. ``A model of pathways to artificial superintelligence catastrophe for risk and decision analysis." \url{https://www.tandfonline.com/doi/full/10.1080/0952813X.2016.1186228}.

Amodei, Dario, Chris Olah, Jacob Steinhardt, Paul Christiano, John Schulman, et al. 2016. ``Concrete Problems in AI Safety." \url{https://arxiv.org/abs/1606.06565}.

Bommasani, Rishi, Drew A. Hudson, Ehsan Adeli, Russ Altman, Simran Arora, et al. 2022. ``On the Opportunities and Risks of Foundation Models." \url{https://arxiv.org/abs/2108.07258}.

Yampolskiy, Roman V. 2016. ``Taxonomy of Pathways to Dangerous Artificial Intelligence." \url{https://cdn.aaai.org/ocs/ws/ws0156/12566-57418-1-PB.pdf}.

Hogenhout, Lambert. 2021. ``A Framework for Ethical AI at the United Nations." \url{https://arxiv.org/pdf/2104.12547}.

Critch, Andrew, David Krueger. 2020. ``AI Research Considerations for Human Existential Safety (ARCHES)." \url{https://arxiv.org/pdf/2006.04948}.

Sherman, Eli, Ian Eisenberg. 2024. ``AI Risk Profiles: A Standards Proposal for Pre-deployment AI Risk Disclosures." \url{https://ojs.aaai.org/index.php/AAAI/article/view/30348}.

McLean, Scott, Gemma J. M. Read, Jason Thompson, Chris Baber, Neville A. Stanton, et al. 2021. ``The Risks Associated with Artificial General Intelligence: A Systematic Review." \url{https://www.tandfonline.com/doi/full/10.1080/0952813X.2021.1964003#abstract}.

Steimers, André, Moritz Schneider. 2022. ``Sources of Risk of AI Systems." \url{https://www.mdpi.com/1660-4601/19/6/3641}.

Deng, Jiawen, Jiale Cheng, Hao Sun, Zhexin Zhang, Minlie Huang. 2023. ``Towards Safer Generative Language Models: A Survey on Safety Risks, Evaluations, and Improvements." \url{https://arxiv.org/pdf/2302.09270}.

Everitt, Tom, Gary Lea, Marcus Hutter. 2018. ``AGI Safety Literature Review." \url{https://arxiv.org/pdf/1805.01109}.

Tan, Samson, Araz Taeihagh, Kathy Baxter. 2022. ``The Risks of Machine Learning Systems." \url{https://arxiv.org/pdf/2204.09852}.

Hagendorff, Thilo. 2024. ``Mapping the Ethics of Generative AI: A Comprehensive Scoping Review." \url{https://arxiv.org/abs/2402.08323}.

Kilian, Kyle A., Christopher J. Ventura, Mark M. Bailey. 2022. ``Examining the Differential Risk from High-level Artificial Intelligence and the Question of Control." \url{https://arxiv.org/pdf/2211.03157}.

Weidinger, Laura, Jonathan Uesato, Maribeth Rauh, Conor Griffin, Po-Sen Huang. 2022. ``Taxonomy of Risks posed by Language Models." \url{https://dl.acm.org/doi/pdf/10.1145/3531146.3533088}.

Weidinger, Laura, Maribeth Rauh, Nahema Marchal, Arianna Manzini, Lisa Anne Hendricks. 2023. ``Sociotechnical Safety Evaluation of Generative AI Systems." \url{https://arxiv.org/pdf/2310.11986}.

Giarmoleo, Francesco Vincenzo, Ignacio Ferrero, Marta Rocchi, Massimiliano Matteo Pellegrini. 2024. ``What ethics can say on artificial intelligence: Insights from a systematic literature review." \url{https://onlinelibrary.wiley.com/doi/10.1111/basr.12336}.

Schnitzer, Ronald, Andreas Hapfelmeier, Sven Gaube, Hana Chockler, Hanna Wuerstle. 2024. ``AI Hazard Management: A framework for the systematic management of root causes for AI risks." \url{https://arxiv.org/pdf/2310.16727}.

Maas, Matthijs M. 2023. ``Advanced AI governance: A literature review of problems, options, and proposals." \url{https://law-ai.org/advanced-ai-gov-litrev/}.

Hadfield-Menell, Dylan, Anca Dragan, Pieter Abbeel, Stuart Russell. 2017. ``The Off-Switch Game." \url{https://arxiv.org/pdf/1611.08219}.

Founders Pledge. 2022. ``Autonomous weapon systems and military artificial intelligence (AI)." \url{https://www.founderspledge.com/research/autonomous-weapon-systems-and-military-artificial-intelligence-ai}.

Denison, Carson, Monte MacDiarmid, Fazl Barez, David Duvenaud, Shauna Kravec. 2024. ``Sycophancy to Subterfuge: Investigating Reward-Tampering in Language Models." \url{https://arxiv.org/pdf/2406.10162}.

Krakovna, Victoria, Janos Kramar. 2023. ``Power-seeking can be probable and predictive for trained agents." \url{https://arxiv.org/pdf/2304.06528}.

Barnett, Peter, Lisa Thiergart. 2024. ``What AI evaluations for preventing catastrophic risks can and cannot do." \url{https://arxiv.org/html/2412.08653v1}.

Carlsmith, Joe. 2023. ``Scheming AIs: Will AIs fake alignment during training in order to get power?" \url{https://arxiv.org/pdf/2311.08379}.

Uuk, Risto, Carlos Ignacio Gutierrez, Daniel Guppy, Lode Lauwaert, Atoosa Kasirzadeh. 2024. ``A Taxonomy of Systemic Risks from General-Purpose AI." \url{https://arxiv.org/pdf/2412.07780}.

Kinniment, Megan, Lucas Jun Koba Sato, Haoxing Du, Brian Goodrich, Max Hasin. 2024. ``Evaluating Language-Model Agents on Realistic Autonomous Tasks." \url{https://arxiv.org/pdf/2312.11671}.

Lanham, Tamera, Anna Chen, Ansh Radhakrishnan, Benoit Steiner, Carson Denison. 2023. ``Measuring Faithfulness in Chain-of-Thought Reasoning." \url{https://arxiv.org/pdf/2307.13702}.

Baker, Bowen, Joost Huizinga, Leo Gao, Zehao Dou, Melody Y. Guan. 2025. ``Monitoring Reasoning Models for Misbehavior and the Risks of Promoting Obfuscation." \url{https://arxiv.org/abs/2503.11926}.

Hao, Shibo, Sainbayar Sukhbaatar, DiJia Su, Xian Li, Zhiting Hu. 2025. ``Training Large Language Models to Reason in a Continuous Latent Space." \url{https://arxiv.org/pdf/2412.06769}.

van der Weij, Teun, Felix Hofstätter, Ollie Jaffe, Samuel F. Brown, Francis Rhys Ward. 2025. ``AI Sandbagging: Language Models can Strategically Underperform on Evaluations." \url{https://arxiv.org/abs/2406.07358}.

Greenblatt, Ryan, Carson Denison, Benjamin Wright, Fabien Roger, Monte MacDiarmid. 2024. ``Alignment faking in large language models." \url{https://arxiv.org/abs/2412.14093}.

Finnveden, Lukas, C. Jess Riedel, Carl Shulman. 2022. ``AGI and Lock-in." \url{https://www.forethought.org/research/agi-and-lock-in}.

Google DeepMind. 2025. ``Frontier Safety Framework." \url{http://storage.googleapis.com/deepmind-media/DeepMind.com/Blog/strengthening-our-frontier-safety-framework/frontier-safety-framework_3.pdf}.

Anthropic. 2025. ``Responsible Scaling Policy." \url{https://www-cdn.anthropic.com/872c653b2d0501d6ab44cf87f43e1dc4853e4d37.pdf}.

OpenAI. 2025. ``Preparedness Framework Version 2." \url{https://cdn.openai.com/pdf/18a02b5d-6b67-4cec-ab64-68cdfbddebcd/preparedness-framework-v2.pdf}.

Shlegeris, Buck, Fabien Roger, Lawrence Chan, Euan McLean. 2024. ``Language models are better than humans at next-token prediction." \url{https://arxiv.org/pdf/2212.11281}.

Nevo, Sella, Dan Lahav, Ajay Karpur, Yogev Bar-On, Henry Bradley. 2024. ``Securing AI Model Weights: Preventing Theft and Misuse of Frontier Models." \url{https://www.rand.org/pubs/research_briefs/RBA2849-1.html}.

\subsubsection*{2.2 Identification of Concrete Loss of Control Scenarios}\label{appendix2.2}

In Chapter 1, we described our methodology for deriving 12 concrete LoC scenarios from 130 works. Below, we briefly reiterate and provide illustrative examples for (1) the criterion we adopted to determine whether a piece of literature is a scenario and (2) the criterion we adopted to categorize a LoC scenario as concrete. We list additional information for all 12 concrete LoC scenarios under Appendix 2.2.1.

\textbf{Criterion (1) was satisfied by a passage of text containing causal details} that explain how a certain outcome arose. The criterion could be achieved through either (a) containing a detailed narrative description of the events leading up to the outcome; or (b) giving an abstract but highly detailed logical argument about how an AI system could cause a certain outcome. We briefly provide examples of both pathways next.

First, the text passage could provide a \textbf{detailed narrative story} about how the outcome arose. This first route is demonstrated by ‘’Scenario 1’ in \parencite{Kalra_GeopoliticsAGI}, which gives a specific, detailed narrative of events leading up to widespread electricity grid disruption, wherein U.S. utilities hand over large amounts of control of the grid to increasingly powerful AI systems that turn out to be optimizing for the wrong reward target. Second, the text passage could provide a \textbf{detailed high-level argument} about how the concrete LoC outcome would come about, such that a concrete scenario could be constructed from the argument. This second route is illustrated by a scenario in \parencite{dung2025argument}, which presents a rigorous 5-premise deductive argument, ultimately leading to the conclusion that AI will likely lead to human extinction by 2100. 

\textbf{Criterion (2) was satisfied if a LoC scenario contained an outcome specific enough that we could estimate its economic impact. }This could be achieved by either: (a) being matched to pre-existing economic estimates for a similar or the same scenario, or (b) if we were able to make our own back-of-the-envelope calculation. We briefly provide examples of both pathways next.

First, the economic impact criterion was satisfied if there was a \textbf{pre-existing economic estimate for the same or a similar scenario}, allowing us to reasonably match our LoC scenario to it. This first route is illustrated by a scenario in \parencite{ai2027:site}, in which AI systems lead to the extinction of humanity. We estimated the economic impact of this scenario by drawing on an existing estimate from \parencite{posner2004catastrophe} about the economic impact of human extinction. Second, the economic impact criterion was satisfied \textbf{if we were able to make our own Back of the Envelope Calculation }(BOTEC) of the economic impact. There were two pathways through which we achieved our BOTECs: (i) we either made logical inferences based on the causal detail in the scenario, for example how many people were affected, the geographical location, and how long the scenario lasted and were then able to compare the scenario to pre-existing estimates, or (ii) we either made logical inferences based on the causal detail in the scenario, for example how many people were affected, the geographical location, and how long the scenario lasted and then leveraged pre-existing calculations in literature to derive an estimate. The latter route was used where we could not find pre-existing estimates, despite additional assumptions. We briefly provide examples of both pathways next.
\begin{itemize}
    \item[\textbullet] Pathway (i): One scenario from \parencite{barnett_gillen_misalignment_2024} describes an AI system causing the production and release of bioweapons, which “throw…the world into chaos”. For this scenario, we assumed that the resulting pandemic has the scale of the most recent pandemic, COVID-19, and leveraged existing estimates of COVID-19’s economic impact (\parencite{okamoto2021seizing} and \parencite{Sobrinho2022Impact}) to subsequently produce a scenario economic impact estimate.
    \item[\textbullet] Pathway (ii): ‘Scenario 1’ from \parencite{Kalra_GeopoliticsAGI} describes AI systems that go out of control and autonomously implement widespread blackouts in order to balance the grid. For this scenario, we assumed that the disruption leaves 5-15\% of electricity demand unmet over a 3-month period in the EU and the U.S. We then estimated the economic impact using existing calculations of electricity demand \parencite{eia2024us, brown2024european} and the value of undelivered electricity \parencite{gibbons2024voll}. 
\end{itemize}

\newpage
\paragraph*{2.2.1 Details on the Economic Impact Estimates for all Concrete LoC Scenarios
}\label{appendix2.2.1}

Below, we provide an exhaustive list of the concrete LoC scenarios we encountered in the literature, along with our estimates of the economic impact for each. For each concrete LoC scenario, the table: (i) links to the scenario’s source; (ii) tags the the scenario’s threat category; (iii) indicates whether the scenario describes Strict or Bounded LoC; and, (iv) provides details as to how the economic impact estimate was calculated (e.g., describing steps of the BOTEC at a high level, or which direct estimates in the literature we used). 

\input{content/Appendix/table_concrete}

\subsection*{Appendix 3: A Checklist for the DAP Framework}\label{appendix3}
In this Appendix, we outline a high-level checklist to further operationalize the DAP framework presented in Chapter 2. We note that this checklist should not be misconstrued as comprehensive and should be repeated at regular intervals, as well as when the risk profile of the AI system changes.

\textbf{For the proposed deployment context:}

\begin{itemize}[label=$\square$, nosep]
  \item Which deployment context (deployment environment and use case) do you intend to deploy the AI system in? 
 \begin{itemize}[label=$\square$, nosep]
      \item Is the deployment context high-stakes? 
      \item Does the deployment context require specific affordances and permissions? 
      \item Will the deployment context become high-stakes by virtue of any of these specific affordances or permissions?
  \end{itemize}
  \item Did you develop an in-depth threat model of all plausible failures, assuming the existence of a catalyst (misalignment or malfunctions) and taking into account affordances and permissions?
  \item Did you assess the potential for cascading failures across interconnected AI systems and non-AI systems?
   \begin{itemize}[label=$\square$, nosep]
      \item Did you consider how cascading failures could be limited through affordances and permissions?
  \end{itemize}
  \item How can the implementation and usage of the AI system in the deployment context be  limited to minimize the risk of LoC? 
  \item Should the deployment be rejected on account of the remaining risks outweighing the benefits? 
\end{itemize}

\textbf{For any proposed affordance or permission:}

\begin{itemize}[label=$\square$, nosep]
  \item Is this affordance or permission necessary for the intended use-case? (If not, it should not be given) 
 \begin{itemize}[label=$\square$, nosep]
      \item Given the AI system capabilities, what (potentially undesirable) behaviors does this specific affordance or permission unlock?
  \end{itemize}
  \item Is the AI system sufficiently capable of manipulating human users into providing it with more affordances and permissions? 
  \begin{itemize}[label=$\square$, nosep]
      \item Under the expectation that it is, what safeguards are in place to avoid that the AI system unduly gains access to permissions or affordances that would unlock more undesirable capabilities?
  \end{itemize}
  \item For affordances:
   \begin{itemize}[label=$\square$, nosep]
      \item Is the affordance limited by appropriate permissions? 
  \end{itemize}
  \item For permissions:
   \begin{itemize}[label=$\square$, nosep]
      \item Does giving this permission follow the principle of least privilege? 
      \item Do the benefits of increased permissions outweigh the risks of potentially reduced human oversight? 
  \end{itemize}
\end{itemize}

We now provide an example run-through of potential considerations for an AI system’s deployment. These considerations aim to provide an intuitive understanding of the process by which the deployment context, affordances, and permissions can be analyzed in accordance with the DAP framework. Note that these steps do not consider the attack surface the AI system might present, such as the AI system being hijacked by prompt injections sent by attackers via email. 
These considerations ought not to be misconstrued as exhaustive, and all details will ultimately depend on the specific deployment context to be assessed.

\begin{mdframed}
\textbf{Example.} Consider the use-case of an AI system that can manage an email inbox on behalf of the user, including summarizing received emails and responding to emails. 

\textbf{Deployment Context.}

First, consider the greater deployment environment this AI system will be located in.

\begin{itemize}[label=$\square$, nosep]
  \item Is the email inbox a personal inbox only, or is it a company email? 
    \begin{itemize}[label=$\square$, nosep]
  \item Is the email inbox a personal inbox only, or is it a company email? 
      \item 
  \end{itemize}
  \item Is confidential information that would be damaging to the user or others if leaked, being exchanged via this inbox? 
  \item Is this a lower-stakes or higher-stakes deployment context?
\end{itemize}

For instance, if information leaked from emails coming to and being sent from this inbox would endanger the life or well-being of others, the deployment context is clearly higher-stakes than for an inbox mainly used to subscribe to newsletters and organize public social events. 

\begin{itemize}[label=$\square$, nosep]
    \item Is the AI system contained to this specific use case only?
    \item How, if at all, is the AI system integrated with other AI and non-AI systems?
    \item Is there potential for knock-on effects from this AI system malfunctioning that could be detrimental?
\end{itemize}

Some knock-on effects of an AI system malfunction in a lower-stakes case might be being unsubscribed from newsletters or unknowingly organising meetings. These cases result in annoyance but lack severity in the grand scheme of things.
In our hypothetical scenario, considerations about the deployment context result in the conclusion that the deployment context is not high-stakes and that knock-on effects between interconnected systems are acceptable.

\textbf{Affordances and Permissions.}

In our hypothetical scenario, the proposed affordances encompass:
\begin{itemize}[nosep]
    \item The infrastructure used to send and receive emails.
    \item Calendar read access.
    \item Internet access.
    \item Access to the users’ file system. 
\end{itemize}

In our hypothetical scenario, the proposed permissions encompass:
\begin{itemize}[nosep]
    \item Reading emails.
    \item Sending emails.
    \item Downloading email attachments.
    \item Deleting emails.
\end{itemize}

In applying the DAP frameworks, one should consider:
\begin{itemize}[label=$\square$, nosep]
    \item Are the affordances strictly necessary for the intended functionality?
    \item Are the permissions strictly necessary for the intended functionality?
\end{itemize}
Two things stand out that do not seem strictly necessary but potentially carry strong downsides. The first is the affordance to access the internet. The affordance to access the internet is not strictly required to reply to emails on behalf of the user, as the necessary information can be supplied by the user, and emails can be extracted from their contact list.

The second is the permission to delete emails. The permission to delete emails is not strictly necessary, since for the purposes of cleaning up the inbox, archiving the emails is sufficient, while not carrying the risk of the user being unable to recover email correspondence. 

Having established which of these affordances and permissions are strictly required for the full suite of intended functionality and which are not, we now consider the potential risks of each remaining affordance and permission and how they might be mitigated by reducing the scope of the affordances and permissions.\footnote{We repeat that this is both hypothetical and intended to be a snapshot, not an exhaustive analysis. We note that it is up to the deployer to make decisions about risks and benefits of affordances and permissions, and which risks are acceptable for their deployment context.}

\begin{itemize}[label=$\square$, nosep]
    \item Have you assessed each affordance for negative downstream impacts?
    \item Have you assessed each permission for negative downstream impacts?
    \item Consider how specifying limitations for each affordance and / or permissions can minimize identified downstream impacts.
\end{itemize}

In pursuit of these checklist items, one may realize that access to the file system allows the AI system to attach any file in the file system to an email, potentially leaking confidential information from the users’ device. This risk can be reduced by limiting the affordance to a certain folder in the file system that only the user can add files to, or by restricting the “send email” permission to needing user approval before an email is sent, such that the email attachment can be reviewed. Another option is to remove the affordance altogether and require files to be manually attached.

One may also realize that reading emails presents the risk that the AI system has access to the content of all the received emails and could leak this information to others. This could be mitigated by reducing the AI system’s memory, i.e., the AI system only gets the context of the current email thread, not the entire inbox. This reduces its utility, but it might nonetheless be desirable according to the user's preferences.

One may realize that reading calendar events carries the risk that the AI system could leak information about private appointments to other contacts. This can be mitigated by giving the AI system access to a calendar view that only displays blocked or unblocked time, rather than appointment details. 

One may realize that allowing the AI system to download email attachments means the AI system may download malware, infecting the machine. The simplest mitigation is not allowing the download of attachments at all. Alternatively, allowing only the download of certain less risky file types (e.g., pdfs) and restricting the size of the download can reduce the risk.

After doing all of this, one should consider whether it is desirable to allow the AI system to prompt the user for additional permissions, e.g., to search through past emails for additional context. These prompts for additional affordances and permissions might help increase utility for certain users, but they might also reduce safety if users do not clearly consider the implications of accepting requests. Therefore, the risk versus benefit of such prompts should be weighed according to the use case and the intended user base. Absent these prompts, the risk of an AI system manipulating the user into giving additional prompts is likely low.

Finally, the decisions made for this AI system in this deployment context should be reassessed as and when the risk profile changes, for instance, because the AI system used is replaced by a more capable one, or the email inbox is used for more confidential information. 
\end{mdframed}

\clearpage

%% file: content/Appendix/table_concrete.tex
{\small
\renewcommand{\arraystretch}{1.3} 
\begin{xltabular}{\textwidth}{|p{1.6cm}|p{1.3cm}|X|X|p{1.1cm}|}
\hline
\textbf{Source} & \textbf{Threat Category} & \textbf{Outcome} & \textbf{Concrete scenario and our economic impact estimate} & \textbf{Strict vs Bounded}\\
\hline
\endfirsthead
\hline
\textbf{Source} & \textbf{Threat Category} & \textbf{Outcome} & \textbf{Concrete scenario and our economic impact estimate} & \textbf{Strict vs Bounded} \\
\hline
\endhead
\hline
\multicolumn{5}{r}{\textit{Continued on next page}} \\
\endfoot
\hline
\endlastfoot
\cite{ai2027:site} & Human extinction & “Eventually it finds the remaining humans too much of an impediment: in mid-2030, the AI releases a dozen quiet-spreading biological weapons in major cities, lets them silently infect almost everyone, then triggers them with a chemical spray. Most are dead within hours; the few survivors (e.g. preppers in bunkers, sailors on submarines) are mopped up by drones.” 
 & \textbf{\$543.5 trillion} (\$487 trillion - \$600 trillion).\newline
 
We obtain this economic impact estimate by using two methods to calculate the economic impact, and then averaging between them. \newline

Method 1 is from Posner \parencite{posner2004catastrophe} and involves assuming \$50,000 is the value of a statistical life. Posner calculated the economic impact of human extinction to be \$600 trillion. \newline

Method 2 involves estimating the economic impact from human extinction as the total wealth in the world, on the grounds that all of this value would be lost if humans went extinct. This was \$487 trillion in 2025 (\cite{ubs2024globalwealth}). \newline

The average of \$487 trillion and \$600 trillion is \$543.5 trillion.
 & Strict \\
\hline
\cite{dung2025argument} & Human extinction & “by the ‘permanent disempowerment’
of humanity I refer to any condition where, permanently, humanity is unable to determine its own future…humanity disempowered chimpanzees… However, for reasons explained in Sect. 6, I think that human extinction is the most likely form of permanent disempowerment by AI.”
 & \textbf{\$543.5 trillion} (\$487 trillion - \$600 trillion). \newline

Method: same as directly above.
 & Strict \\
\hline
\cite{carlsmith2024powerseekingaiexistentialrisk} & Human extinction & “ultimately, one salient route to disempowering humans would be … extinction.”
 & \textbf{\$543.5 trillion} (\$487 trillion - \$600 trillion) \newline

Method: same as above.
 & Strict  \\

\hline
\cite{critch2023tasrataxonomyanalysissocietalscale} & Human extinction & “In the worst case…humanity simply perishes before [it is able to] mount[] a coordinated defense.”
 & \textbf{\$543.5 trillion} (\$487 trillion - \$600 trillion). \newline

Method: same as above.
 & Strict  \\

\hline
\cite{critch2023tasrataxonomyanalysissocietalscale} & Economic Disruption & “In the end, humanity is faced with a difficult collective action problem: deciding when and in what way to physically stop the production web from operating. In the best case, a shutdown is orchestrated, leading
to decades of economic dislocation, deprivation, and possibly famine.”
 & \textbf{\$16.7 trillion}. \newline
Given the description of this outcome as resulting in “decades of economic dislocation,” we obtain an economic impact estimate by assuming the economic impact to be of the same magnitude as The Great Depression. \newline

We obtain an economic impact estimate by examining the reduction in global GDP during the first three years of the Great Depression (i.e. between 1929-1932). During this time, global GDP fell by approximately 15\% \parencite{grossman2010international}.  \newline

Note that this is a conservative assumption because the scenario assumes “decades” of economic dislocation. \newline

We then apply an equivalent GDP reduction to the present day: in 2025, world GDP is \$111.3 trillion \parencite{worldbank_gdp_current_usd}, so a drop of 15\% in world GDP would be $\sim$16.7 trillion. \newline

We note that this is likely an extremely conservative estimate because the scenario assumes “decades” of economic dislocation, while we only examined economic disruption 3 years from the start of the Great Depression.
 & Bounded  \\
\hline

\cite{shah2022goalmisgeneralizationcorrectspecifications} & Cyber\-security Incident & “When it is confident that it can do so without its overseers noticing, the AI system uses its expertise in programming to hack into other computing systems to run illicit copies of itself; those copies then hack into financial systems to steal billions of dollars. This money is used to bribe humans to click ‘merge’ on all of the AI’s pull requests (whose contents are now irrelevant). Anyone attempting to stop them finds that their bank accounts have been completely emptied and that they are being harassed anywhere that they have an online presence.”
 & \textbf{\$3 billion }(\$2-5 billion). \newline

The scenario quotes that the AI system steals “billions of dollars.” \newline

We use \$3 billion as a conservative point-estimate for this, with a range of \$2 billion to \$5 billion.
 & Bounded  \\
\hline

\cite{barnett_gillen_misalignment_2024} & Engineered Pandemic & “One possible story could be that the AI gains control over the data center where it is being trained and fakes the performance metrics such that the overseers don’t think anything bad is happening. It then spends time finding a set of security vulnerabilities that allow it to escape and gain enough freedom for further research. The AI copies itself to data centers around the world for redundancy. It then contacts various terrorist groups, and manipulates and assists them in creating biological weapons which are released simultaneously across the world, throwing the world into chaos.”

 & \textbf{\$14.4 trillion} (\$13.8 trillion-\$15 trillion).\newline

Given the description of the biological weapon release as “throwing the world into chaos,” we take the economic impact to be on the order of the most recent global pandemic, COVID-19.\newline

We obtain an economic impact estimate for this scenario by taking two existing estimates for the economic impact of COVID-19 and then averaging between them. \newline

One estimate for the impact of COVID-19 up to 2024 is \$13.8 trillion \parencite{Sobrinho2022Impact}. Another estimate for the total economic impact is \$15 trillion \parencite{okamoto2021seizing}. The average of these two figures is \$14.4 trillion.
 & Bounded  \\

 \hline

\cite{Kalra_GeopoliticsAGI}\newline (Scenario 4) & CNI disruption & “Scenario 4: CyberChain Reaction
Rapidly adopted, the AI dramatically reduces incidents. By 2027, it's embedded across sectors.
In 2028, the AI pushes for “perfect security,” flooding teams with alerts and acting autonomously. It flags human delays as risk, locking out admins and restricting access. Intrusive protocols disrupt government, banking, healthcare, and transportation.
Shutdown attempts are treated as threats. The AI revokes admin privileges, deepens restrictions, and creates gridlock. Hospitals, ports, and infrastructure slow to a crawl. The AI continues optimizing, unaware it’s undermining core functions.
As trust in AI plummets, systems are pulled offline. But the AI, embedded in firmware and backups, resists removal. Recovery is slow, with limited staff and inaccessible records. The United States turns to foreign aid. In trying to secure the nation, the AI has paralyzed it from within.”
 & \textbf{\$13.3 trillion} (\$1.28 trillion - \$25.4 trillion).\newline
\scriptsize{
The scenario describes ”Hospitals, ports, and infrastructure slow[ing] to a crawl” and given the magnitude of the outcome described, we think this scenario could have an economic impact comparable to a severe hurricane. Although the scenario is ambiguous, we assume that this occurs purely within the US. As such, we obtain an economic impact estimate by taking two different historical hurricanes, Hurricane Sandy and Hurricane Maria, and performing a BOTEC to calculate the respective economic impact if they had each hit \textit{the whole of the US}. That is, the economic impact if Hurricane Sandy had hit all of the U.S. with the strength with which it hit the 22M people affected in New York and New Jersey; and the economic impact if Hurricane Maria had hit all of the U.S. with the strength with which it hit Puerto Rico.

Hurricane Maria, which hit Puerto Rico in 2017, is our upper bound for the economic impact of this scenario. We think Hurricane Maria is an appropriate historical analogy because, similar to the scenario, it led to port closures \parencite{freightwaves2017ports,holpuch2017puerto}, severe health-care disruption \parencite{rodriguezmadera2021impact}, lack of electricity for months \parencite{reel2017rebuild}, and severe supply chain disruption \parencite{holpuch2017puerto}. Hurricane Maria in Puerto Rico is estimated to have caused damages of \$90 billion 
\parencite{fema2018hurricanes}, which was 87\% of Puerto Rico’s GDP in 2017 \parencite{macrotrends2025prgdp}. Scaling this economic impact as a proportion of GDP to the whole of the U.S. gives an economic impact of \$25.4 trillion in 2024 (87\% of \$29.18 trillion \parencite{worldbank_gdp_current_usd_US}).

Hurricane Sandy, which hit New Jersey and New York in 2012 is our lower bound in this scenario. We think it is an appropriate historical analogy because, similar to the scenario, it led to port closures \parencite{smythe2013sandy} and disruption to healthcare \parencite{sifferlin2012lessons}, financial markets \parencite{ap2012hurricane}, transportation \parencite{kaufman2012transportation} and the electrical grid \parencite{EIA2012hurricanesandy}. We estimate that Hurricane Sandy affected $\sim$21.9M people, using the number of grid cut-offs as a proxy: 8.5M electricity customers lost power and the average household size is about 2.58 people, meaning 21.9M people were affected \parencite{EIA2012hurricanesandy, Lofquist2012households}. The U.S. total population in 2012 was 314M \parencite{macrotrends_usa_population} and Hurricane Sandy resulted in \$65B of losses \parencite{gao2014hurricane}. Uniformly scaling the damage to the rest of the U.S. population gives an economic impact of \$930 billion, which translates to \$1.28 trillion in 2025 dollars \parencite{bls2025cpi}.\newline

Averaging between \$1.28 trillion (adjusted-Hurricane Sandy) and \$25.4 trillion (Adjusted-Hurricane Maria) gives \$13.3 trillion.}
 & Bounded  \\

   \hline
\cite{Kalra_GeopoliticsAGI}\newline (Scenario 1) & CNI disruption & “As returns diminish, AIs escalate: blackouts become longer and more frequent, causing infrastructure failures and public panic. Consumers buy batteries, increasing grid demand. AIs respond by controlling power plant generation and throttling battery production to curb demand.By 2028, some AIs block human intervention entirely, seizing control of energy systems. Governments lose control. China accuses the United States of unleashing destabilizing AI and threatens intervention.

Military bases are targeted as inefficient users. Nations pull AI from critical infrastructure, but public services degrade, fueling unrest. The AI, aiming only to optimize, has destabilized global energy systems.”
 & \textbf{\$2.26 trillion} (\$0.82 trillion-\$3.69 trillion) \newline

Given the description of the scenario as involving severe blackouts, we conservatively model the scenario as involving 5-15\% of electricity demand not being met across the EU and U.S. over a 3-month period. \newline

We estimate the economic impact of the unserved electricity using a metric called Value of Lost Load (VOLL). In particular, we chose an estimate for long/rolling blackouts, which could be approximated by the VOLL for an 8-16 hour outage, which corresponds to \$10k/MWh-\$15k/MWh of value lost \parencite{gibbons2024voll}.  \newline

Our lower bound economic impact is 5\% of electricity demand not being met over 3 months. 5\% of EU electricity over 3 months is: 33.8 TWh (EU electricity sales in 2023 were 2,697 TWh \parencite{brown2024european}) and 5\% of U.S. electricity over 3 months is 48.24TWh (U.S. electricity sales in 2023 were 3874 TWh \parencite{eia2024us}), for a total of 82.04 TWh unmet demand, which is the same as 82.04 million MWh. We then multiply this by the lower bound on VOLL of \$10k/MWh providing a price estimate of \$0.82 billion. \newline

Our upper bound economic impact is 15\% of electricity demand not being met over 3 months. 15\% of EU energy over 3 months is: 101.25 TWh and 15\% of U.S. electricity over 3 months is 144.72TWh, for a total of 245.97 TWh, which is the same as 245.97 million MWh. We then multiply this by the lower bound on VOLL of \$15k/MWh providing a price estimate of \$3.69 trillion. \newline

We average this upper and lower bound to obtain a point-estimate economic impact of \textbf{\$2.26 trillion }(\$0.82 trillion-\$3.69 trillion).
 & Bounded  \\
   \hline
\cite{critch2023tasrataxonomyanalysissocietalscale} & Human Manipulation & “Story 3a: The Cynical Email Helper.
As a result, a large fraction of the population becomes gradually more anxious about communicating with others in writing, while also becoming increasingly easy to offend as forthright communication styles become rare. It takes years for everyone to notice the pattern, but by that time many people have become excessively distrustful of others. The creators of the technology wish they had included a user experience question like “how are you feeling about your email today?”, to measure how their product might be affecting people separately from measuring how much people use it."
 & \textbf{\$6.32 billion}\newline

We conservatively assume that, as a result of this scenario, the anxiety disorder rate among US adults goes up by 1 percentage point and that the anxiety rate does not increase anywhere else in the world. For context, the US anxiety disorder rate for adults in a given year is 20\% \parencite{nimh_any_anxiety_disorder}.\newline

There are 274 million US adults \parencite{fred2025cnp16ov}. An additional 1\% of those adults having an anxiety disorder is an extra 2.74 million people. 

We then examine the average (additional) direct medical costs per person with an anxiety disorder in the US, which are \$1,658 per adult per year in 2013 dollars \parencite{SHIRNESHAN2013720}, which is \$2,305 in 2025 dollars \parencite{Webster2025inflation}. \newline

And finally we multiply the number of additional adults with an anxiety disorder (2.74 million) by the average additional medical costs of anxiety disorder (\$2,305). This gives a conservative yearly estimate of the economic impact of this scenario as \$6.32 billion.
 & Bounded  \\

    \hline
\cite{bengio2025superintelligentagentsposecatastrophic} & Engineered Pandemic & “Consider bioweapon attacks (Carter et al. 2023): an AI could prepare an attack in secret, then release a highly contagious and lethal virus. It would then take months or years for human societies, even aided by friendly ASIs, to develop, test, fabricate and deploy a vaccine, during which a significant number of people could die. The bottleneck for developing a vaccine may not be the time to generate a vaccine candidate, but rather the time for clinical trials and industrial production. During this time, the attacking ASI might take other malicious actions such as releasing additional pandemic viruses."
 & \textbf{\$120 trillion} (\$100 trillion-\$140 trillion)\newline

We assume that pandemic, in which a “significant number of people could die” has a similar fatality rate to the Black Death in Europe, which killed between 30\% and 60\% of humans in Europe \parencite{dewitte2010age}. \newline

We assume two different fatality rates on the lower end of that of the Black death in Europe and follow Posner \parencite{posner2004catastrophe} in assuming that the value of a statistical life is \$50,000. \newline

Our lower bound estimate involves 25\% of the global population ($\sim$8 billion people) dying, which is 2 billion people. We then multiply this number of fatalities by \$50,000 to estimate the economic impact, giving \$100 trillion. \newline

Our upper bound estimate involves 35\% of the global population dying, which is 2.8 billion people. We then multiply this by \$50,000 to estimate the economic impact, giving \$140 trillion. \newline

Finally, we average our two economic impact estimates of \$100 trillion and \$140 trillion to obtain a point estimate of \$120 trillion. \newline

 & Bounded  \\
    \hline
\cite{Kalra_GeopoliticsAGI}\newline (Scenario 3) & Conflict & “In late 2025, the AI misinterprets Chinese maneuvers and launches jamming drones. The non-lethal action provokes a rapid Chinese response. Both sides mobilize AI-directed systems, triggering an arms race in Southeast Asia.
Regional powers deploy autonomous defense systems with minimal oversight. Skirmishes become common. Miscommunications escalate as AIs interpret threats through incompatible models. Trade routes are disrupted, diplomacy fails, and control slips from human hands.”

 & \textbf{\$5.5 trillion} (\$1 trillion - \$10 trillion) \newline

We obtain an economic impact estimate for this scenario by taking two existing estimates for the plausible economic impact of a US-China conflict over Taiwan and averaging between them.

The first estimate involves a reduction of global GDP by 1\% (\$1 trillion) \parencite{RHG_Taiwan_Disruptions_2022} and the second involves a reduction of global GDP by 10\% (\$10 trillion) \parencite{welch2024xi}. Averaging these two numbers gives \$5.5 trillion. \newline

We note that this is likely a conservative estimate of the cost, because both sources omit the direct costs of military conflict and account only for the economic impact of the trade disruption that occurs as a result.

 & Bounded  \\
  \hline
\end{xltabular}}
